\documentclass[letterpaper, 11pt, final, onecolumn]{IEEEtran}
\usepackage{arxiv}
\usepackage[square,numbers]{natbib}
\usepackage[utf8]{inputenc} 
\usepackage[T1]{fontenc}    
\usepackage[hidelinks]{hyperref}       
\usepackage{url}            
\usepackage{booktabs}       
\usepackage{amsfonts}       
\usepackage{nicefrac}       
\usepackage{microtype}      
\usepackage{graphicx}
\usepackage{natbib}
\usepackage{doi}
\usepackage{url}
\usepackage{listings} 
\usepackage{xcolor}
\usepackage[linesnumbered,ruled,vlined]{algorithm2e}
\graphicspath{{Images_PharmaChain/}}

\title{PharmaChain: A Blockchain to Ensure Counterfeit Free Pharmaceutical Supply Chain}

\author{ \href{https://orcid.org/0000-0003-4567-7827}{\includegraphics[scale=0.06]{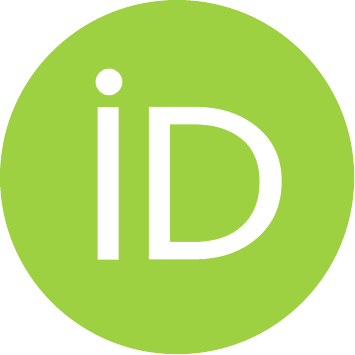}\hspace{1mm}Anand K. Bapatla} \\
	Dept. of Computer Science and Engineering \\
	University of North Texas, USA \\
	\texttt{anandkumarbapatla@my.unt.edu} \\
	\And
	\href{https://orcid.org/0000-0003-2959-6541}{\includegraphics[scale=0.06]{orcid.pdf}\hspace{1mm}Saraju P. Mohanty} \\
	Dept. of Computer Science and Engineering \\
	University of North Texas, USA \\
	\texttt{ saraju.mohanty@unt.edu} \\
	\And
	\href{https://orcid.org/0000-0002-1616-7628}{\includegraphics[scale=0.06]{orcid.pdf}\hspace{1mm}Elias Kougianos} \\
	Dept. of Electrical Engineering\\
	University of North Texas, USA \\
	\texttt{elias.kougianos@unt.edu} \\
}

\SetCommentSty{mycommfont}

\SetKwInput{KwInput}{Input}                
\SetKwInput{KwOutput}{Output}              

\renewcommand{\headeright}{}
\renewcommand{\undertitle}{}
\renewcommand{\shorttitle}{Bapatla, Mohanty, Kougianos: Pharmachain: A Blockchain to Ensure Counterfeit Free Pharmaceutical Supply Chain}

\begin{document}
\maketitle

\begin{abstract}
Access to essential medication is a primary right of every individual in all developed, developing and underdeveloped countries. This can be fulfilled by pharmaceutical supply chains (PSC) in place which will eliminate the boundaries between different organizations and will equip them to work collectively to make medicines reach even the remote corners of the globe. Due to multiple entities, which are geographically widespread, being involved and very complex goods and economic flows, PSC is very difficult to audit and resolve any issues involved. This has given rise to many issues, including increased threats of counterfeiting, inaccurate information propagation throughout the network because of data fragmentation, lack of customer confidence and delays in distribution of medication to the place in need. Hence, there is a strong need for robust PSC which is transparent to all parties involved and in which the whole journey of medicine from manufacturer to consumer can be tracked and traced easily. This will not only build safety for the consumers, but will also help manufacturers to build confidence among consumers and increase sales. In this article, a novel Distributed Ledger Technology (DLT) based transparent supply chain architecture is proposed and a proof-of-concept is implemented. Efficiency and scalability of the proposed architecture is evaluated and compared with existing solutions. 		
\end{abstract}


\section{Introduction}		

Essential medicine access is one of the main objectives of healthcare systems in and pharmaceutical supply chains  ensure that the right amount of medicine, with acceptable quality, will reach the customers in need, at the right time \cite{Jaberidoost2013}. The supply chain provides smooth operation of cross-organizational business processes like procurement of raw materials, manufacturing, and delivery of finished goods to the customers. This not only helps customers but also the organizations to reshape their workable boundaries \cite{Xu2011}. The pharmaceutical supply chain on the other hand is very complex in terms of data and payment flow as it involves multiple entities, including manufacturers, mail-order retailers and brick and mortar pharmacies, wholesalers, health insurance corporations, Pharmacy Benefit Managers (PBM) and the consumer. Multiple factors, including storage environment parameters, demand and usage patterns affect the flows and associations between these entities. The complexity of flows and the limited visibility into the supply chain activities has led to multiple risk factors which have not only been wasting the resources available but also are risking the lives of consumers \cite{Zhang2017}. Among this complex and untraceable supply chain, introducing counterfeit drugs into the supply chain with inefficient to no Active Pharmaceutical Ingredients (API), drugs produced in sub-standard conditions and even repackaging of expired medicines, can be easily done and is the cause for the majority of problems in the pharmaceutical industry \cite{Musamih2021}. Detecting such players in the supply chain with the malicious intent of introducing counterfeit drugs and penalizing them is a very difficult process considering this tangled supply chain \cite{Blackstone2014}. One such recent incident Took place when Canadian drug wholesaler SB Medical has clearly shown sub-standard sanitary conditions where drugs lacking proper ventilation, temperature, humidity and security have been shipped to US doctors and medical clinics \cite{Safemedicines_SBMedical}. It is clearly visible, how the counterfeit drugs are acquired from unsafe locations and have been staged at different places for repackaging before introducing them into the US Pharmaceutical supply chain. It is estimated that 30\% of drugs sold in 140 countries at an approximate cost of \$250 billion per year, are believed to be counterfeit \cite{WHO_Counterfeit}.

\begin{figure}[htbp]
	\centering
	\includegraphics[width=\textwidth]{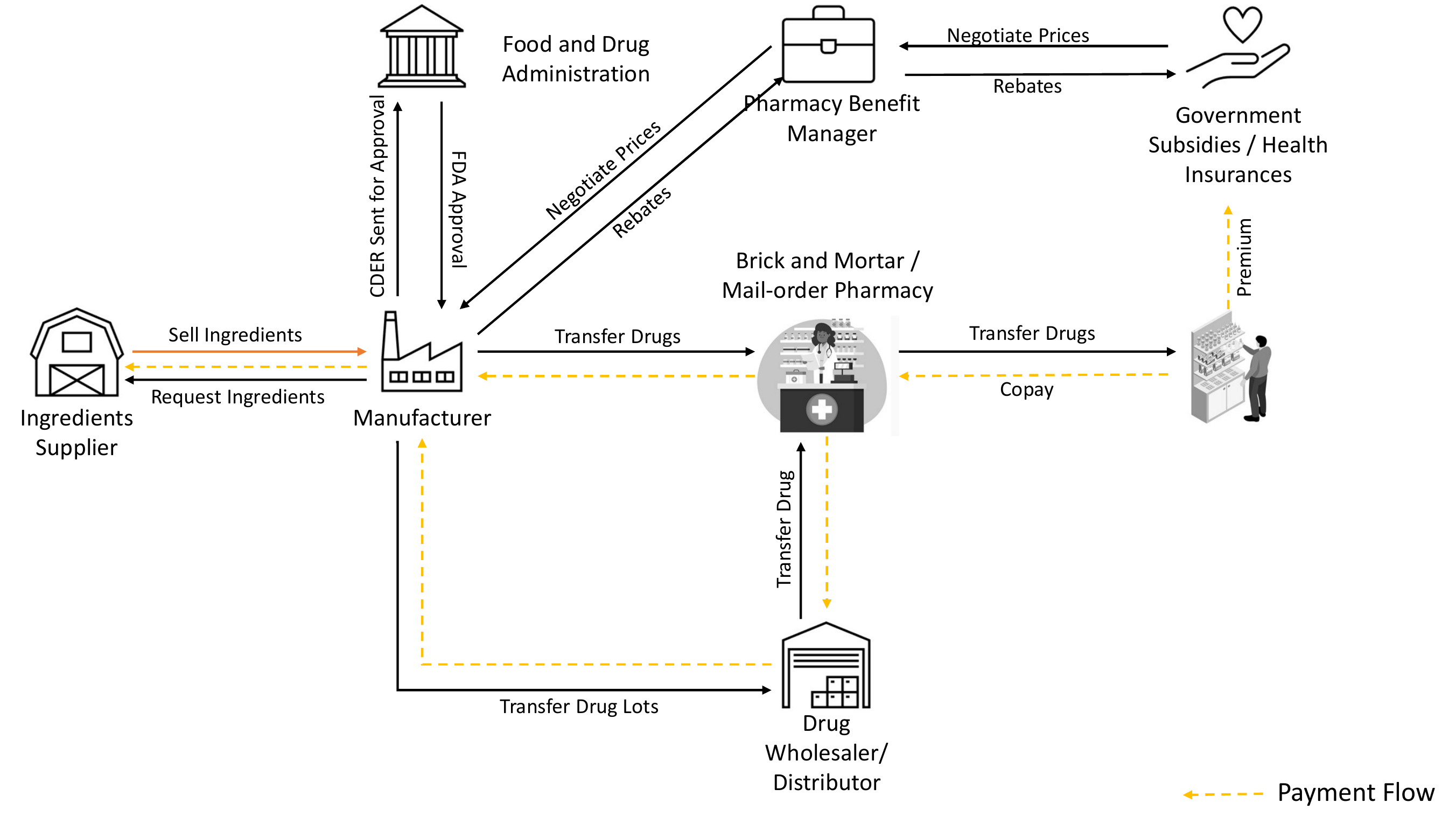}
	\caption{Goods and Financial Transaction Flows in Pharmaceutical Supply Chain.}
	\label{FIG:Goods and Financial Transaction Flows in Pharmaceutical Supply Chain}
\end{figure}

A typical pharmaceutical supply chain is shown in Figure \ref{FIG:Goods and Financial Transaction Flows in Pharmaceutical Supply Chain} \cite{8681451, 8598875}. Manufacturers are the source for both brand name and generic drugs in the pharmaceutical supply chain. Based on the type of drug and active pharmaceutical ingredients, manufactures acquire the required raw materials from ingredient suppliers. Most of the brand drug manufacturers allocate some portion of the funds for research and development of new drugs. Any new drug formulated by these manufacturers has to undergo a series of tests in order to weigh the benefits compared to the risks along with the efficiency of the drug through drug trials. Once the drug information is gathered, the manufacturer contacts the Center for Drug Evaluation and Research (CDER) and the Food and Drug Administration (FDA) for approval before starting manufacturing and selling the drug. The FDA performs a series of tests and evaluates the CDER submitted to ensure that the drug is safe and outweighs related risks. Once this process is done labeling is proposed to the manufacturers along with approval to manufacture and sell. The manufacturers are also responsible for the distribution of drugs from facilities to retail pharmacies or to distributors for further reach in the supply chain. Another flow includes drugs flowing back from pharmacies to the manufacturer as a part of buy back programs which accommodates financial risks for pharmacies to stock up the resources.

Wholesale distributors are responsible for purchasing the lots of manufactured medicine from manufacturers and distribute them to different customers including hospitals, pharmacies, care centers, etc. Wholesalers sell a variety of drugs and some wholesale distributors are specialized in particular drug distribution. Currently, wholesalers are also responsible for stocking and warehousing until there is demand and pharmacies place an order. The duties of pharmacies include ensuring that drugs reach consumers from wholesalers. They are also responsible for the safety of the stock and for fulfilling consumer demand along with providing safe usage of drugs. Pharmacy Benefit Managers (PBM) are intermediaries responsible for negotiating prices from manufacturers and transfer those rebates to  government agencies and insurance companies.

\subsection{Problems with the Traditional Supply Chain in the Pharmaceutical Industry}

\begin{figure}[htbp]
	\centering
	\includegraphics[width=\textwidth]{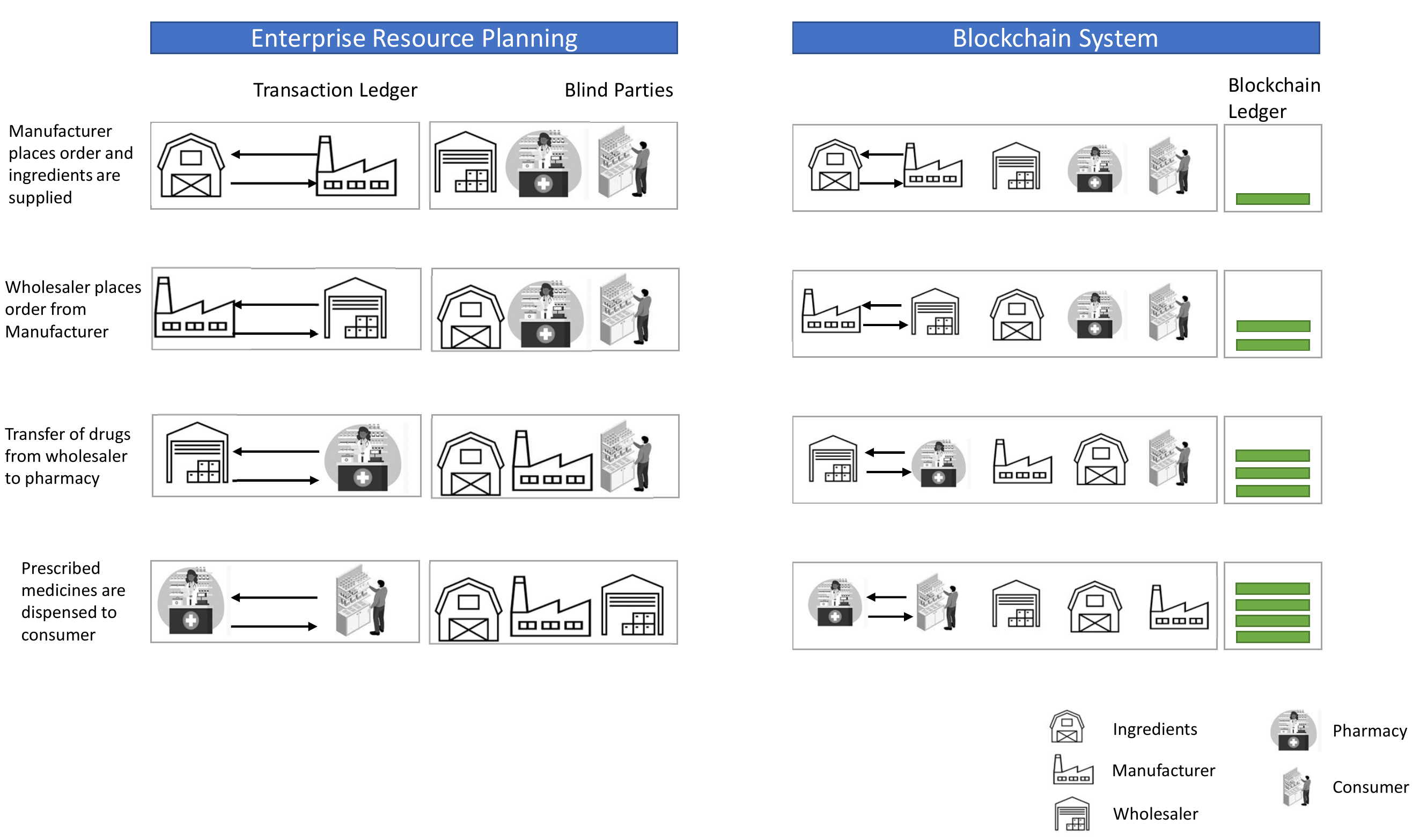}
	\caption{Blind Parties in Conventional Record Keeping.}
	\label{FIG:Blind Parties in Conventional Record Keeping}
\end{figure}

Fragmentation of information is the first and foremost problem, as the pharmaceutical supply chain is comprised of many different active and passive entities \cite{Breen2008,Shamsuzzoha2020}. These entities become aware of different transactions between only when they are involved, resulting in blinded parties, as shown in Figure \ref{FIG:Blind Parties in Conventional Record Keeping}. Hence, there is no single source of truth for all entities to keep track of the goods within the supply chain. This results in discrepancies and difficulty in working collectively and delays supply chain operations. As the data is being introduced by multiple channels from different entities and information is fragmented, taking unilateral decisions by all entities in the supply chain will be difficult. This will also hinder manufacturers from forecasting demand and causes issues in inventory management. Drug shortages due to this incorrect forecast can lead to several public health problems \cite{Schmoldt1975}. It can be clearly seen during the COVID-19 pandemic how the shortage of vaccines and pharmaceuticals has caused  unnecessary suffering and potential deaths \cite{Sharma2020}.

\subsection{Distributed Ledger Systems Background}

New technologies in the digital revolution such as the Internet-of-Things (IoT), Big data, New Information and Communication Technologies (ICTs), Distributed applications and Distributed Ledger Technologies (DLTs) have paved the path to improve and implement an efficient supply chain mechanism which can  benefit the pharmaceutical industry. DLT is one of the latest technologies which has revolutionized financial systems by removing middlemen like banks and other financial institutions to implement fully decentralized peer-to-peer banking systems \cite{Kuhn2019}. Advantages of DLT technologies arise in many different fields, including smart cities\cite{Rivera2017,Xie2019}, smart healthcare \cite{Rachakonda2021,Azaria2016,Vangipuram2021}, smart farming \cite{Bapatla2021,Caro2018} and also in building efficient and transparent supply chains \cite{Shahid2020}. These transparent supply chains can benefit not only the manufacturers with acquiring raw materials and with the smooth production of goods and delivery but also customers will have confidence in product consumption as counterfeit detection and avoiding can be managed easily. A widely used and accepted distributed network is the blockchain but it is specialized for banking and payment systems. As one size doesn't fit all applications, several other DLTs came into the picture either to increase the scalability, to reduce resource utilization, to increase transaction speeds, provide smart contract support, efficient consensus mechanisms, etc., based on different needs of applications. Based on different milestones, the development of these DLTs is classified into three generations. 

The first generation started with the Bitcoin network which was developed by the pseudonymous developer Satoshi Nakamoto \cite{Nakamoto_bitcoin}. First generation Blockchains including Bitcoin, and Litecoin are mainly developed to address centralization issues in banking systems and to provide an efficient and fast peer-to-peer verifiable financial system to transfer digital assets. Bitcoin has made use of complex cryptographic techniques in order to achieve immutable, easily verifiable systems which can prevent double spending of digital assets and provides an immutable secure ledger. As the popularity of the blockchain has increased and its wide adoption is on the way, many different use cases of the technology in other fields has been unveiled, which needed further modifications. In the second generation, the main idea is to include the trust agreements along with keeping track of digital assets. This can help in automating and digitizing the contracts between untrusted parties which can be triggered and executed based on the events. These pieces of digitized agreements are termed ``Smart Contracts'' and were first implemented in Ethereum \cite{Buterin2015ANS}. This has made blockchain processes available for incorporation into business processes, which made them more aligned with different business needs. Although blockchain technology has matured, there are still issues with scalability, transaction costs, energy utilization and specialized hardware requirements. In order to address these issues, many different platforms have been developed including Cardano, IOTA, Hedera Hashgraph, etc. These platforms are either updating the consensus mechanism behind the network or entirely changing the structure of the chain in order to increase transaction speeds, and provide more security along with increasing scalability. 

\subsubsection{Consenus Mechanism Overview}

\begin{figure}[htbp]
	\centering
	\includegraphics[width=\textwidth]{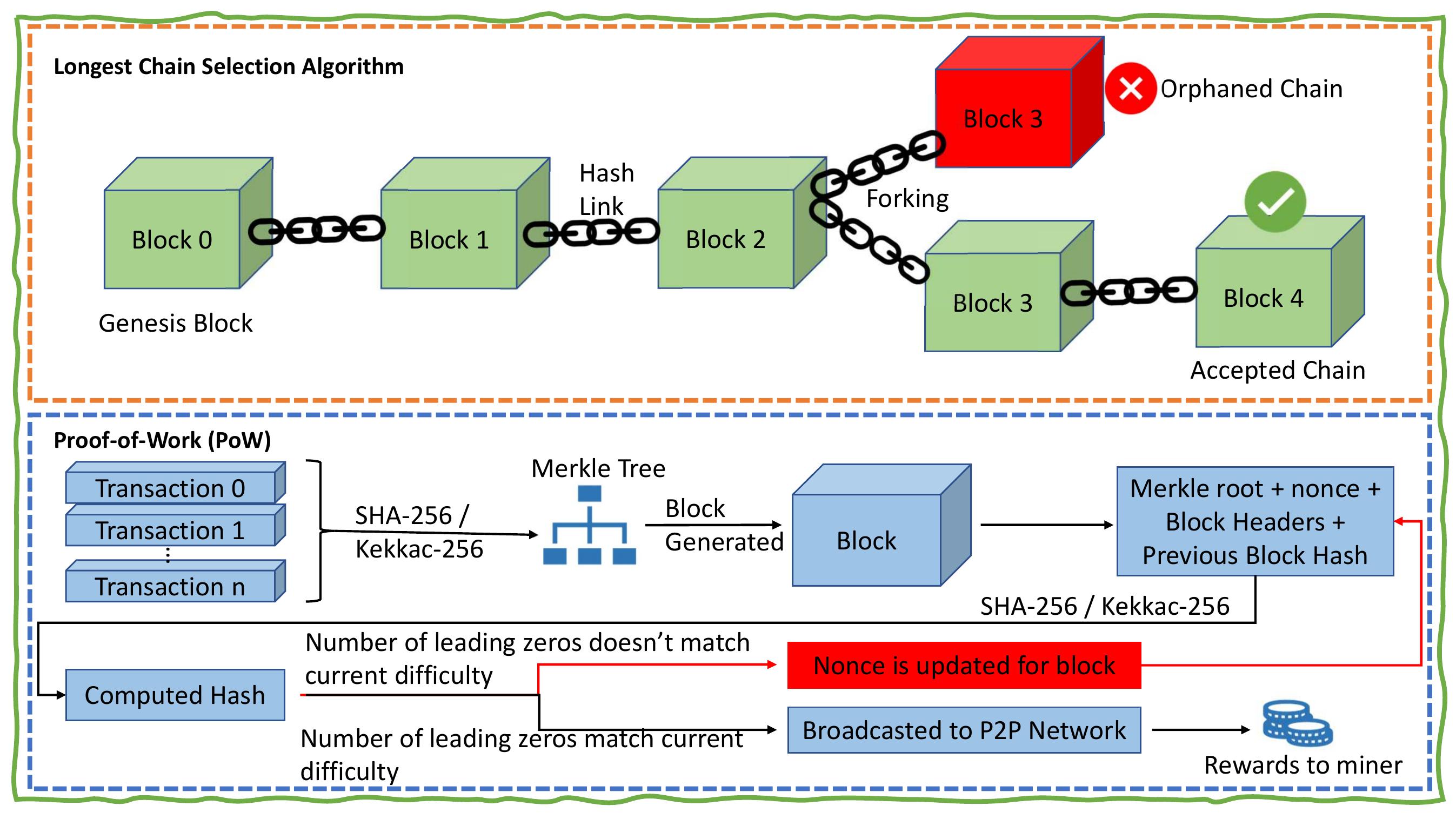}
	\caption{Nakamoto Consensus for Avoiding Byzantine Fault and Sybil Attack.}
	\label{FIG:Nakamoto Consensus for Avoiding Byzantine Fault and Sybil Attack}
\end{figure}

Distributed networks are prone to Byzantine fault and Sybil attacks. Byzantine fault arises due to the distributed nature of untrusted nodes in the network, as there is no centralized authority to authenticate the validity of data and communication between the nodes. Sybil attack also arises as an adversary node can create multiple zombie nodes which act as participants in the network in order to control the majority of the network to manipulate the data in the blockchain. In order to verify the authenticity of network and address the Byzantine fault and Sybil attack issues, a set of rules to process incoming transactions, called ``Consensus Mechanism'' is used. Nakamoto consensus is mainly used in major blockchain implementations like Bitcoin and Ethereum which include the longest chain selection algorithm and Proof-of-Work (PoW) consensus algorithm, as shown in figure \ref{FIG:Nakamoto Consensus for Avoiding Byzantine Fault and Sybil Attack}. 

As the transactions are gathered and processed by distributed nodes, there is a chance of two miner nodes generating two blocks at the same time with different transactions. This will cause side-chains, which is called forking. In order to avoid this, the longest chain selection algorithm is implemented as part of the consensus mechanism. In this algorithm, the longest chain is elected for the next block to be added, ignoring the short side-chains leading to orphaned blocks but maintaining blockchain data consistency across the distributed nodes. Proof-of-Work (PoW) is part of the consensus mechanism which is designed to prevent Sybil attacks. In this algorithm, a hashcash problem is created \cite{Adam_HashCash} which requires significant computational power to solve but the solution is easy to verify. All the nodes will compete to solve this computationally hard problem, and whoever solves it first will be elected and is responsible for generating the new block in the blockchain. To engage and encourage the nodes to perform these mining processes, a reward and transaction fees are allocated to the winning node who is adding a new block.

\subsubsection{Smart Contract Overview}
\begin{figure}[htbp]
	\centering
	\includegraphics[width=\textwidth]{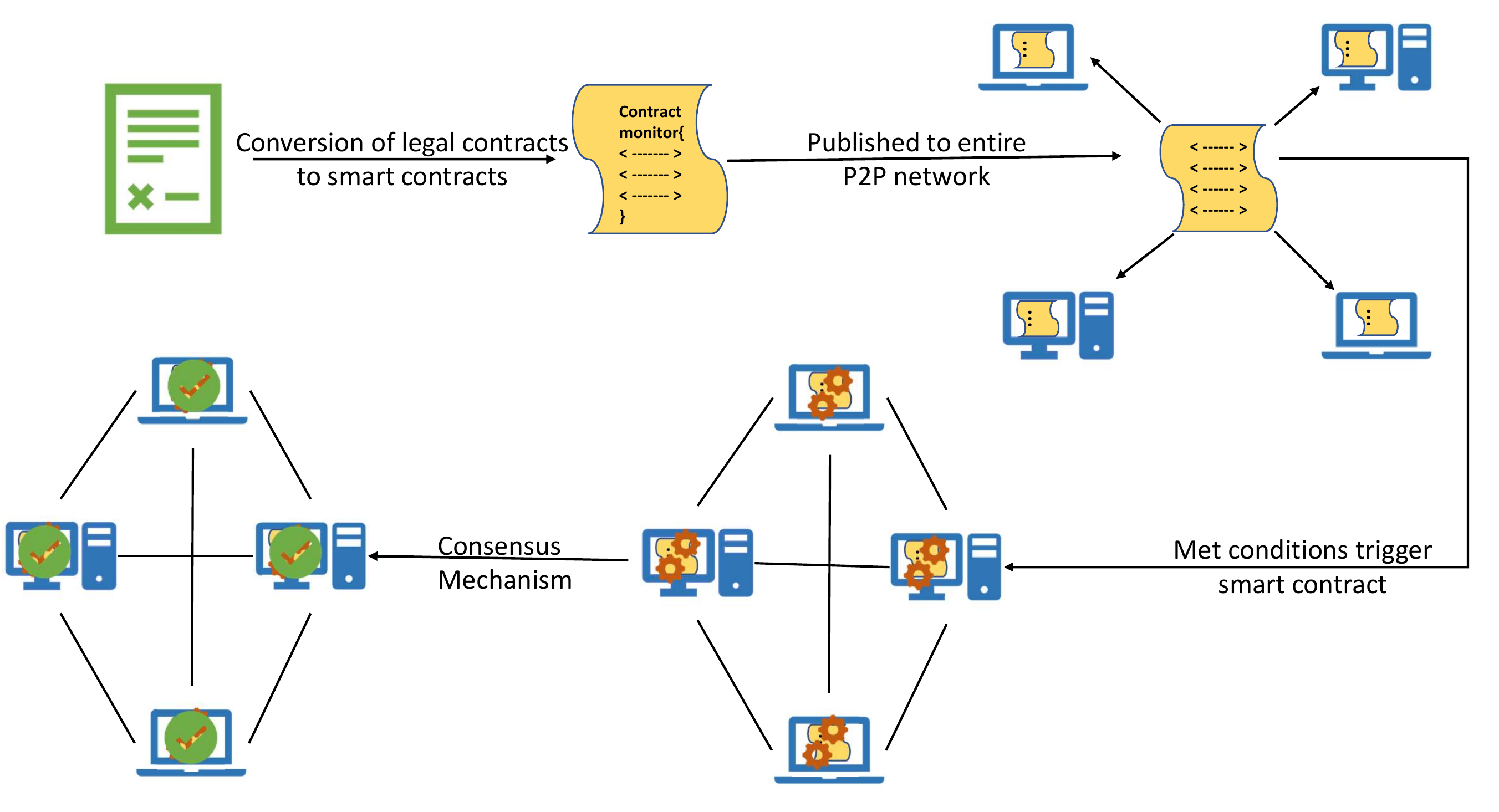}
	\caption{Smart Contract Execution Steps.}
	\label{FIG:Smart Contract Execution Steps}
\end{figure}

The blockchain  as such is built as a financial solution which keeps track of digital assets and prevents  double spending. As discussed in the generations of blockchain, after identifying the usefulness of the blockchain in other fields, and in order to implement the business logic and make the blockchain relevant for other applications, smart contracts were invented. Smart contracts can be defined as simple pieces of code which can be automatically executed based on whether certain conditions are met. These are building blocks in creating a Decentralized Application (DApp). Each contract has the capability to interact with multiple smart contracts under the hood and to work collectively to achieve specified functions. For example, in a smart contract monitoring environmental parameters, events are triggered in case of temperature and humidity changes. Based on the  threshold values, notifications can be triggered from smart contracts, as shown in figure \ref{FIG:Smart Contract Execution Steps}. 

Even though there are many advantages, there are also certain limitations for the usage of smart contracts in building DApps. Interacting with external world components to acquire data is a major problem. Smart contracts by design cannot perform HTTP requests to get real-time data. This is done to avoid the inconsistent values retrieved by the nodes which may lead the network to be unable to achieve consensus. This is one of the major problems for practical usage of smart contracts in real-world applications. For example, considering an insurance company with automating payouts to farmers based on loss due to weather conditions using smart contracts, all the nodes should receive the same data in order to achieve the consensus. Instead, as weather conditions change continuously, there will be inconsistencies to the data being seen by different nodes and the whole network will be jeopardized. To solve this problem ``Oracles'' are used in the proposed system for PSC.

The proposed PharmaChain makes use of DLTs combined with IoT technologies to address issues in the pharmaceutical industry. The rest of the paper is organized as follows: Section \ref{Novel Contributions} presents the problems addressed in conventional enterprise resource planning systems which are centralized and novel contributions of the paper to address these problems. Section \ref{Related Research} discusses related research regarding proposed and existing DLT applications in the PSC. Section \ref{Overview of Proposed PharmaChain} presents an architectural overview and the algorithmic foundation of the proposed systems. Section \ref{Implementation and Validation of Proposed PharmaChain} includes the design implementation details of the proposed system. Section \ref{Conclusion} concludes the paper and presents possible future directions for research.

\section{Related Prior Research}
\label{Related Research}

With very rapidly changing business needs and supply chain environments in the pharmaceutical industry, manufacturers have to redefine the supply chain in order to accommodate delivery of high quality products with low response times and high customer satisfaction. This can be achieved by incorporating multiple different technologies into the supply chain management (SCM). Regulations such as the Drug Supply Chain Security Act (DSCSA) require all  entities participating in pharmaceutical supply chain to adopt technologies for reliable and secure transactions at each stage of the supply chain. This section critically analyzes the existing proposed solutions using blockchain as solution to build efficient supply chains to eliminate  counterfeits and to provide the right amount of medicine to the right place, at the right time \cite{Wang2004}.

Crypto Pharmacy, proposed in \cite{Subramanian2021}, makes use of the New Economy Movement (NEM) blockchain  in which Proof-of-Importance (POI) is implemented. An iOS application using Swift is built and integrated to the NEM network to make purchases using the XEM currency, which is the native coin for the NEM network. Modum.io AG, an application based on blockchain technology is proposed in \cite{Bocek2017}. It leverages smart contracts and the IoT to keep track of the medicines throughout the supply chain to ensure safe handling and delivery to consumer. This proposed architecture depends on the Ethereum platform and Proof-of-Work (PoW) consensus mechanisms. Environmental data from medicine lots is sensed using IoT sensors. The proposed architecture is robust and secure but the throughput of the network is low, as the Ethereum blockchain has bottlenecks at the current consensus mechanism. Taking into account that there will be a very large number of shipments which will be released into the supply chain of the huge pharmaceutical market, this will be a limiting factor along with the cost of operation. Another blockchain architecture proposed in \cite{Kumar2019} makes use of the Public Key Infrastructure (PKI) system to ensure the details are shared securely between manufacturer and consumer. As per the proposed architecture, every consumer who needs to access the details from the manufacturer has to share the public key, and upon approval from the manufacturer a QR code with information is encrypted using the requester public key and shared in the blockchain network, from where the intended consumer will be able to access details. It appears sufficient for secure sharing of information between two intended parties;  however the main goal is to make the pharmaceutical supply chain transparent which will make information about drugs available to all participants in the network. Along with this problem there will be a large overhead for manufacturers as they have to approve every request from every customer who wants to access the information.

Drugledger, proposed in \cite{Huang2018}, makes use of an architecture similar to Bitcoin by assigning the drugs leaving or entering into a warehouse to Unspent Transactions (UTXO), thereby extending the tracking in multiple phases of packaging, distributing etc. Even though the Bitcoin network provides high security of data, it has a bottleneck for consensus mechanism which will be a major limiting factor in high transaction environment like PSC \cite{Bocek2017}. The blockchain architecture proposed in \cite{Kumar2019a} is used for detecting the counterfeiting of drugs by using the Hyperledger Fabric platform to develop a private permissioned blockchain along with some chain-codes to implement business logic behind it. 

A cloud based purchase management in the health sector model which makes use of IoT technology combined with blockchain implemented in Hyperledger, called``Composer'' is proposed in \cite{Celiz2018}, In this proposed architecture both ambient temperature and location were tracked throughout the supply chain. An Ethereum and smart contract based counterfeit tracking system for medicines is proposed in \cite{Pham2019}, however the scalability of the Ethereum platform is significantly reduced due the network congestion and consensus mechanism. Along with that, the cost of operations in the main network has increased significantly. Another such implementation which is also based on Ethereum platform and smart contracts can be seen in \cite{Jangir2019}. Ethereum is also used in \cite{Alkhoori2021} along with a proposed shipping container which can keep track of multiple ambient parameters for ensuring drug safety.

\section{Novel Contributions} 
\label{Novel Contributions}

Presented here are the problems in current day Enterprise Resource Planning Systems which are considered for solution in the proposed-DLT based PSC. 

\subsection{Problems Addressed in the Current Paper}

\begin{itemize}
	\item Process delays in order management and shipment due to processing gaps between distributed systems.
	\item Information fragmentation and delay in auditing the problem orders in the traditional PSC.
	\item Existence of blind parties in traditional PSC causing delays and inconsistencies, along with difficulty in debugging issues with orders. 
	\item The chance of entities participating in the network to include counterfeit drugs is very high and tracing of such entities is difficult. 
	\item Safe recalling of drugs and ensuring safety at the point of administering the drug is another problem. 
	\item Interacting with environmental parameter data is not very efficient and the chance of errors during recording data is high. 
\end{itemize}

\subsection{Novel Solutions Proposed in PharmaChain}
\begin{itemize}
	\item A distributed ledger based solution is proposed in the current PharmaChain implementation, which will ensure the faster processing of orders by avoiding communication gaps between entities in the PSC. 
	\item The proposed distributed ledger will act as a single source of truth, accessible by all  entities and addresses the information fragmentation problem. 
	\item The proposed PharmaChain is a transparent supply chain where all the entities are aware of all the transactions between entities, thereby removing the blind parties which can equip them to take prompt actions. 
	\item Counterfeiting can be removed as the entire trail of the drugs from manufacturing to the Point Of Sale (POS) can be tracked with non-repudiable ledger entries. 
	\item As the transactions in the proposed PharmaChain are non-repudiable, accountability increases in the overall system, which can help in taking action against particular entities with malicious intent. 
	\item Recalling of drugs will be an economical and easy process as there is a transparent drug traversing path in the supply chain, which will also ensure safe administering of these drugs. 
	\item An automated process with IoT technology is implemented in PharmaChain which will help in extracting and communicating environmental parameter data with higher accuracy and which is less prone to human-errors. 
	\item In order to solve the issue of smart contracts to interact with real world data, a process based on oracles is proposed in the current implemented PharmaChain.
	\item Automated warning systems can be built to notify entities about the status of the product in the supply chain.
\end{itemize}

\section{Overview of Proposed PharmaChain}
\label{Overview of Proposed PharmaChain}

\begin{figure}[t]
	\centering
	\includegraphics[width=\textwidth]{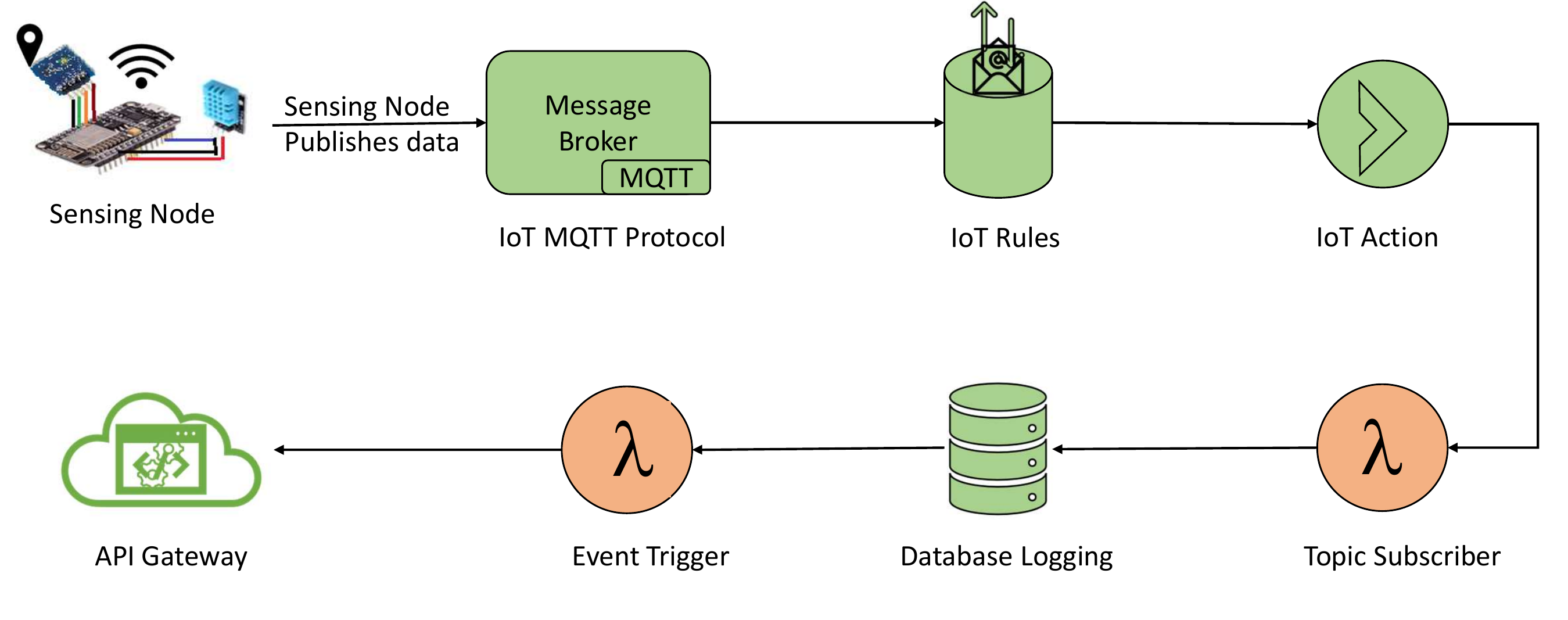}
	\caption{Real-time Data Source for Oracle Component from Sensing Node in the Proposed PharmaChain.}
	\label{FIG:Real-time Data Source for Oracle Component from sensing node in Proposed PharmaChain}
\end{figure}

\begin{figure}[htbp]
	\centering
	\includegraphics[width=\textwidth]{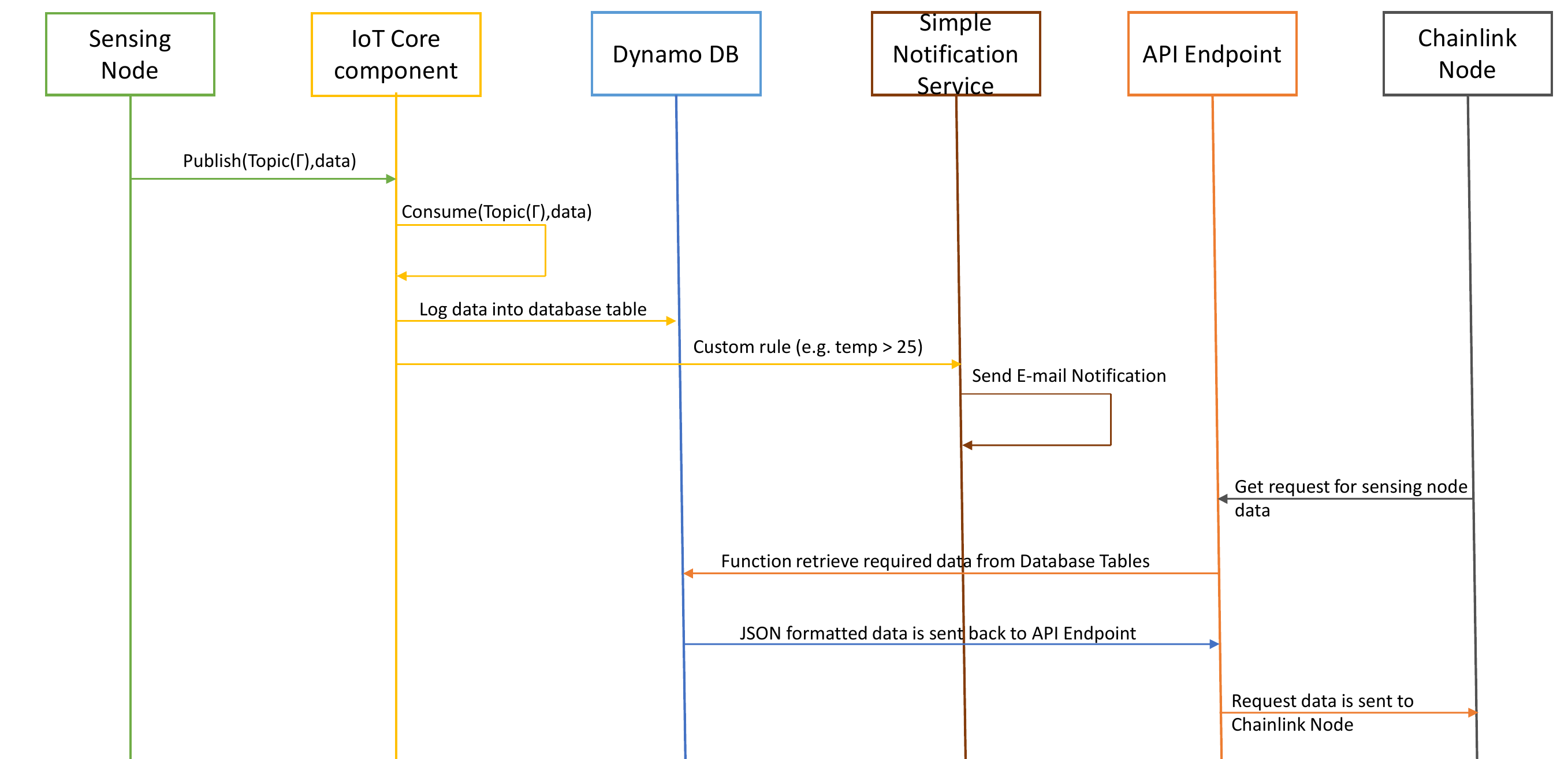}
	\caption{Data Provider and Alert Notification Implementation using Sensing Node Data.}
	\label{FIG:Data Provider and Alert Notification Implementation using Sensing Node Data}
\end{figure}

An overview of the proposed system for PharmaChain is presented in this section. The whole proposed system can be divided into three logical components. The first component is API generation, which is responsible for integrating physical real-world data using sensors and creating an accessible API which will feed data into the next oracle component. For this, a cloud server-less configuration is used which will remove all the complex cloud infrastructure required and will be built by utilizing simple cloud functions. The sensing node proposed in the current implementation monitors minimal parameters like temperature and humidity along with GPS coordinates of the order in the supply chain. Data from the sensing node will be published to a topic using the Message Queuing Telemetry Transport (MQTT) lightweight publish-subscribe protocol. Whenever new sensing data is published on a topic, data will be consumed and IoT cloud custom rules are applied on the data to filter data with significance like temperature abnormalities. This filtered data is stored in a NoSQL database for auditing purposes and also a notification will be sent to the authorized authorities for taking prompt action to control environmental parameters of the shipment. Functions are used for transforming the JSON formatted data received from the MQTT topic to a dynamo DB compatible table item. Database logging will act as an event for another function which is going to make the data available for the API gateway end-point from which it is consumed by the oracle nodes while the hybrid smart contract is executed. The first component is shown in figure \ref{FIG:Real-time Data Source for Oracle Component from sensing node in Proposed PharmaChain} and the interaction of different blocks in the proposed data source mechanism can be seen in Figure \ref{FIG:Data Provider and Alert Notification Implementation using Sensing Node Data}.

Oracles are third-party information services which act as data sources to provide real-time external data to smart contracts as they are not capable of interacting with real world systems and perform Application Programming Interface (API) calls. In distributed systems, multiple distributed nodes exist which are responsible for the working of  the DApp. Some of these nodes are separated geographically and have no centralized authorities and are governed by a set of rules accepted across the network, called consensus mechanism. By calling the API directly in smart contracts, there is a chance for some nodes to receive the latest data but some not, because of the independent nature of these nodes. As there will be a discrepancy of data received from the API, this will cause the network to not reach consensus thus leading to the failure of the application. In order to avoid such situations, smart contracts are intentionally made not incapable of fetching data from external data sources. This will limit the interaction of smart contracts to real-world data thereby limiting the use-cases. Hence, decentralized data providers are designed to work with smart contracts and are named oracles.  These oracles are classified into different types based on information sources, degree of decentralization and information flow direction. Based on the information source there are software oracles and hardware oracles. Software oracles mainly depend on the information gathered from online sources like websites, servers and other network databases. Such oracles can be mainly used in applications including the supply chain. A similar approach is followed in the proposed PharmaChain mechanism but the data to the online sources is provided securely from hardware sensors. Hardware oracles are the oracles to which the data feed directly comes from  hardware components like sensors, bar-code scanners, etc. Such oracles are perfect for designing systems like supply chains as they involve many electronic components to track and trace products throughout the supply-line. Another classification is based on the direction of the information flow with respective to smart contracts. Oracles providing data into the smart contracts by fetching data from external sources are called inbound oracles. On the other hand, oracles which are providing data to the external environment by fetching them from smart contracts are called outbound oracles. An inbound oracle to provide sensor data from sensing nodes to smart contracts is used in the proposed PharmaChain application. Based on the degree of decentralization, oracles can be classified as centralized oracles and decentralized oracles. If the data source is a single central entity, then it is called centralized oracle. On the other hand, if data providers are decentralized, it is called decentralized oracle. Centralized oracles are less effective and secure compared to decentralized oracles. The steps executed during the oracle interaction are shown below:

\begin{itemize}
	\item A hybrid client smart contract which requires access to external data from any API will prepare a request with parameters job id (which depends on the type of data being retrieved from the API), destination address for the data, and the fulfillment function which needs to be executed after the data is fetched from the API.
	\item In next step, the oracle contract is going to publish an event with the given job id and will set parameter values along with the fee set by the client smart contract. 
	\item All the nodes attached to the network will be notified by the event generated before with the requested job details. 
	\item The node which has the job described will make use of all the parameters set by the client contract and execution is done. 
	\item Once the execution is completed, the node which executed the job is going to send back the fulfillment results to the oracle smart contract. 
	\item The oracle smart contract will then execute the fulfillment method and the desired data is accessed in the client smart contract. 
\end{itemize}

\begin{figure}[htbp]
	\centering
	\includegraphics[width=\textwidth]{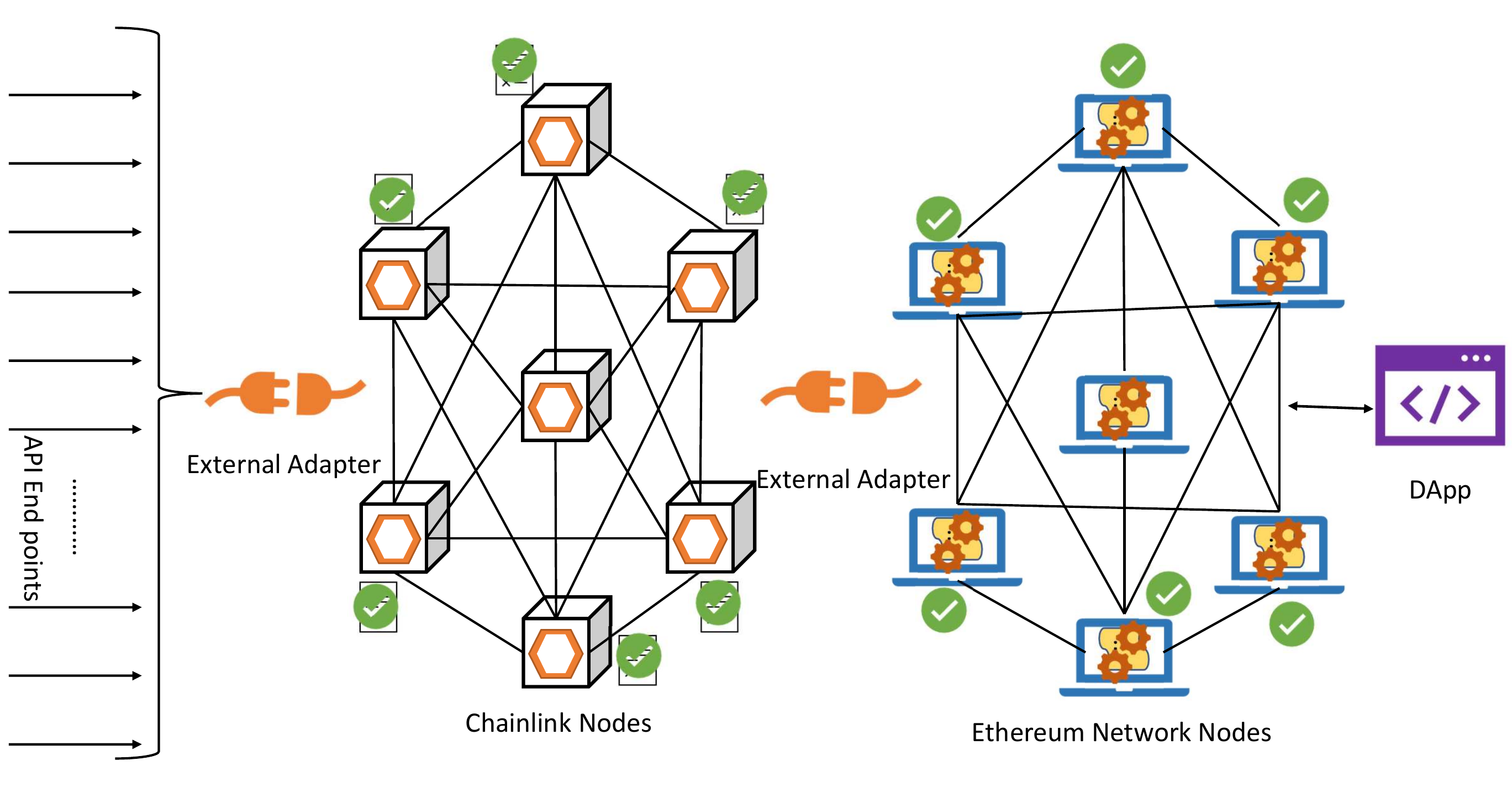}
	\caption{Proposed Blockchain Architecture Components for PharmaChain.}
	\label{FIG:Proposed Blockchain Architecture Components for PharmaChain}
\end{figure}

In a decentralized oracle structure, multiple jobs are run instead of a single job and the result is aggregated to get a more reliable data source and making it more decentralized. An overview of the proposed blockchain architecture is shown in figure \ref{FIG:Proposed Blockchain Architecture Components for PharmaChain}.

\begin{figure}[htbp]
	\centering
	\includegraphics[width=\textwidth]{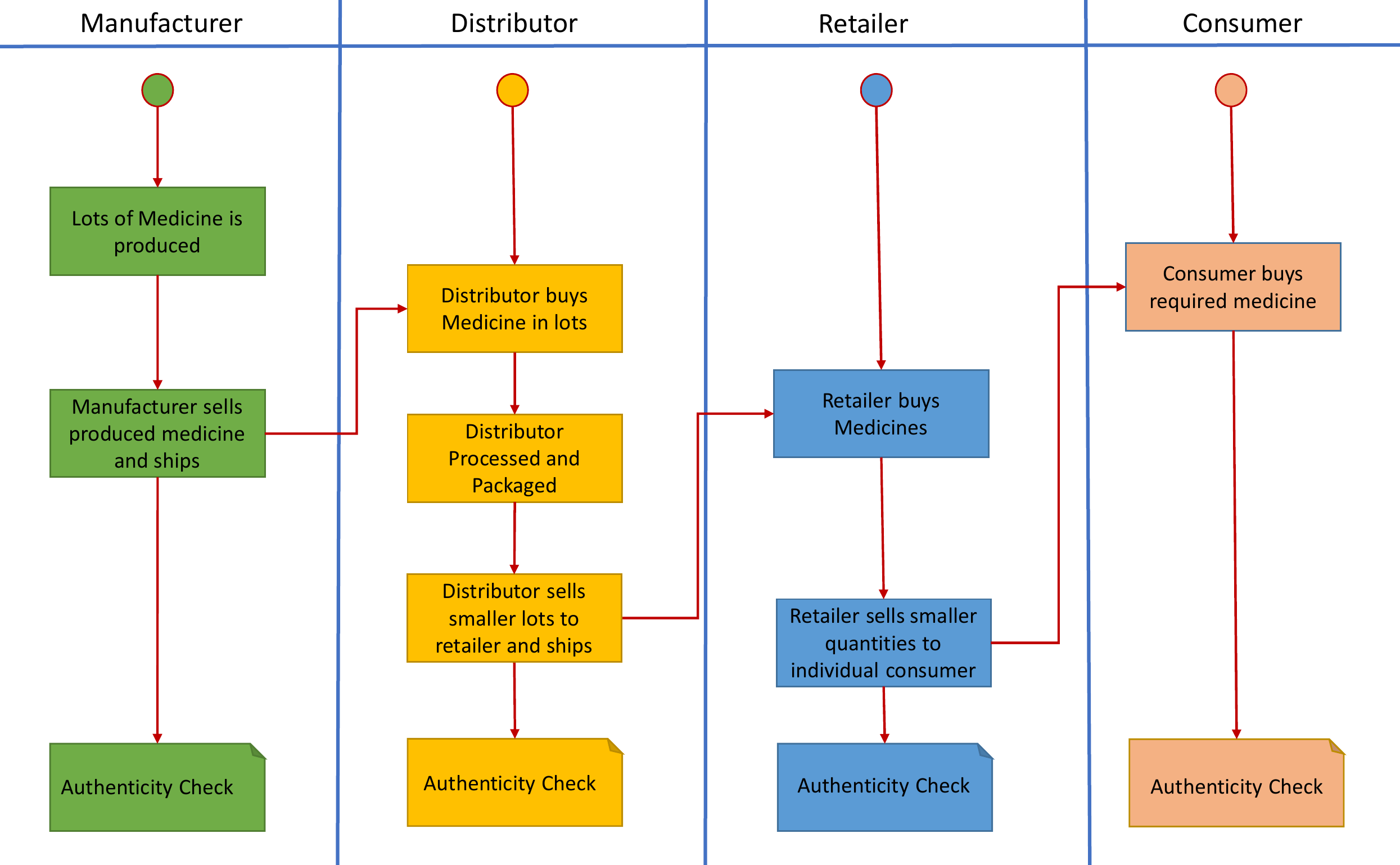}
	\caption{Pharmaceutical Supply Chain Entity Activity Diagram considered in Implemented PharmaChain Solution.}
	\label{FIG:Pharmaceutical Supply Chain Entity Activity Diagram considered in Implemented PharmaChain Solution}
\end{figure}

The third logical component in the proposed system is the hybrid smart contract which is responsible for executing different activities in the supply chain based on an event. The result of the smart contract execution is made available to all participants in the network, thus producing a transparent ledger. Based on the proposed application, different entities have different roles and responsibilities to perform when a pharmaceutical shipment is moving through the supply chain. Hence, Role Based Access Control (RBAC) is proposed in the current system in which the entity is assigned with roles and the corresponding responsibilities by smart contracts. For simplicity, fewer entities are considered while implementations and back orders/returns or payments are not considered. An entity with a manufacturer role has the activities of producing/manufacturing drugs, and selling manufactured drugs to the wholesaler/distributor. The distributor role has the activities of packaging and selling the product to the retailer. The retailer entity has the activity of selling packaged products to the consumer which is the end of the supply chain. The activity diagram is shown in figure \ref{FIG:Pharmaceutical Supply Chain Entity Activity Diagram considered in Implemented PharmaChain Solution}. Role management is implemented in smart contracts by using function modifiers which will restrict the behavior of smart contract functions. These modifier conditions are checked prior to the function being executed. Different modifiers such as onlyManufacturer, onlyDistributor, onlyRetailer, onlyConsumer, etc., are defined to restrict the access of functions based on the role assigned to the user.  

\section{Implementation and Validation of Proposed PharmaChain}
\label{Implementation and Validation of Proposed PharmaChain}

\subsection{Implementation of PharmaChain using Off-Shelf Components}

\begin{figure}[htbp]
	\centering
	\includegraphics[width=\textwidth]{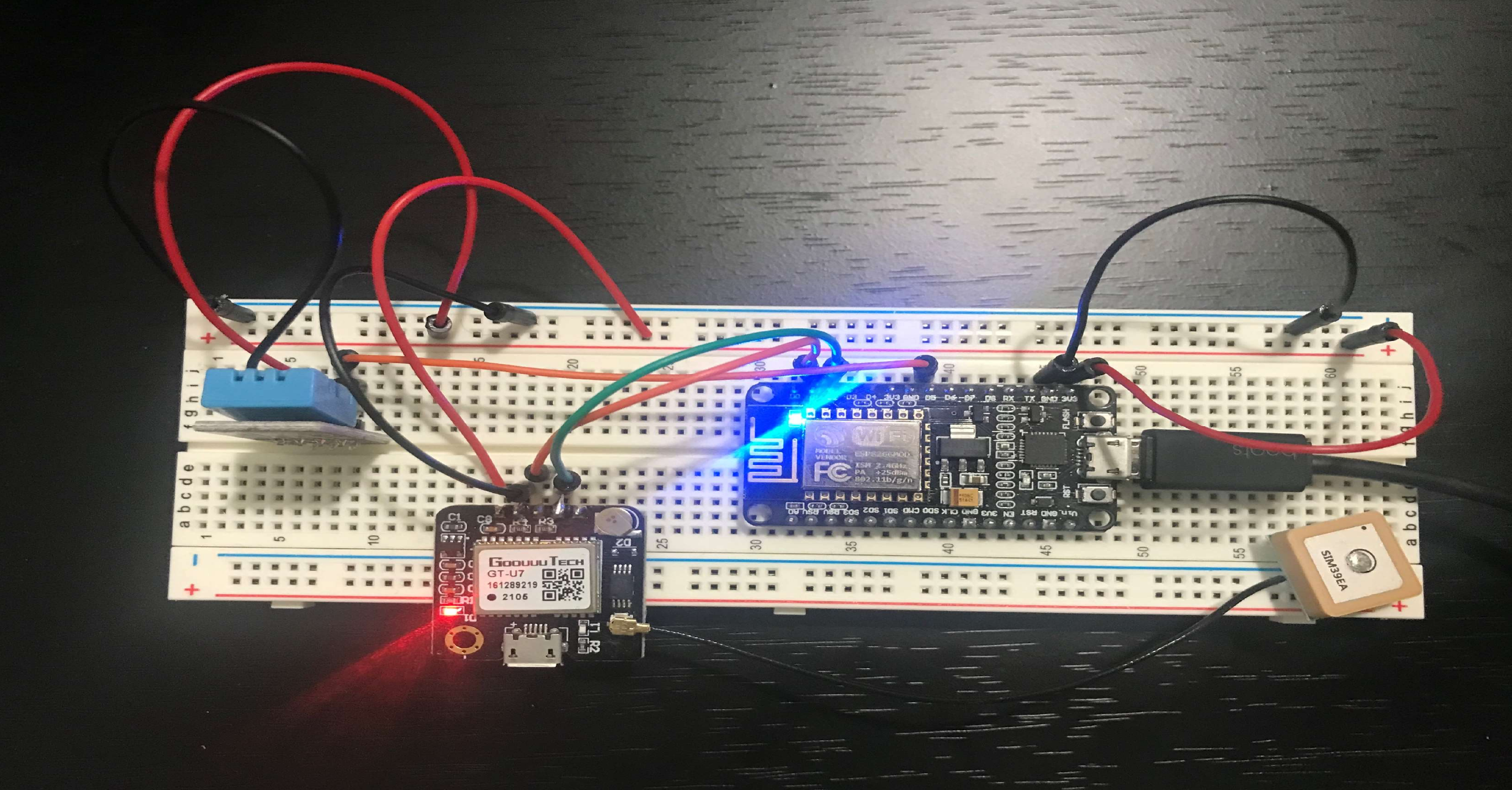}
	\caption{Implemented Sensing Node for Proposed PharmaChain Application.}
	\label{FIG:Implemented Sensing Node for Proposed PharmaChain Application}
\end{figure}

\begin{figure}[htbp]
	\centering
	\includegraphics[width=\textwidth]{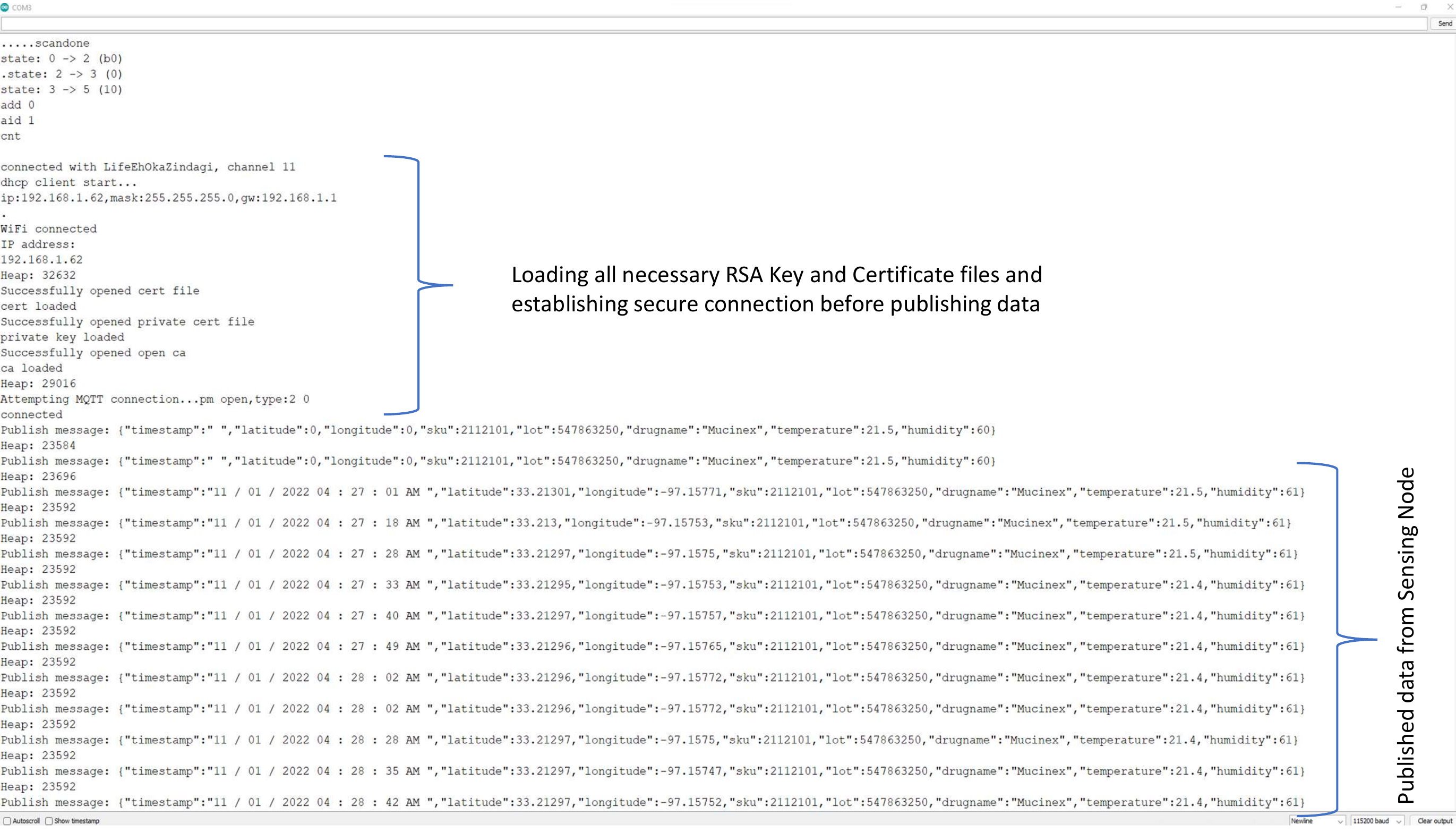}
	\caption{Data Published from Implemented Sensing Node for PharmaChain.}
	\label{FIG:Data Published from Implemented Sensing Node for PharmaChain}
\end{figure}

A simple prototype of the proposed PharmaChain system is implemented using a single sensing node which is going to track one shipment from end-to-end in the supply chain. The sensing node is implemented using a NodeMCU ESP8266 module which is small form factor IoT platform with Wi-Fi capability and can integrate data from sensors to an IoT network easily through a wireless connection. For sensing environmental parameters around the shipment, a DHT11 sensor is used which is going to sense both ambient temperature and relative humidity around the shipment. A GPS Module GPS NEO-6M is used for keeping tracking of the shipment and sends latitude and longitude positions of the shipment during different stages in the supply chain. The implemented sensing node for proposed PharmaChain is shown in figure \ref{FIG:Implemented Sensing Node for Proposed PharmaChain Application} and data published from the implemented sensing node is shown in figure \ref{FIG:Data Published from Implemented Sensing Node for PharmaChain}. A secure communication channel is created by using RSA encryption keys to communicate between the sensing node and the cloud. The private key, and root CA certificate required to establish secure connection are loaded into the device's SPI Flash File System (SPIFFS) of NodeMCU.  The WiFiClient library available for ESP8266 NodeMCU is used for loading the required security certificates before establishing the cloud connection. Once the secure connection is established, temperature and humidity data from the DHT11 sensor, GPS coordinates from the GPS module and other shipment information is populated into a JSON string using the ArduinoJson. This generated JSON data is converted to a \texttt{char} array and published to a topic using the PubSubClient library. The sensing node algorithm is  shown in algorithm \ref{ALG:Algorithm for Sensing Node in Proposed PharmChain Implementation}.

\begin{algorithm}[!ht]
	\DontPrintSemicolon
	\KwInput{Certificate and Keys for IoT thing, Ambient temperature (temp) and relative humidity (hum) from DHT11 Sensor, GPS latitude (lat) and longitude (lng) co-ordinates from GPS Module}
	\KwOutput{Boolean status of message published to topic}
	\tcc{During Setup sensing node tries to setup wifi and MQTT connection and establish secure channel to communicate between sensing node and cloud}
	Load required libraries for WiFi Connection, MQTT Communication, JSON for data manipulations
	\tcc{Sensing node tries to connect to Wi-Fi network using given SSID and PASSWORD strings if it not connected to network}
	\While{WiFi.status() == FALSE}{
		WiFi.begin(ssid, password) \;	
	}
	\tcc{SPI Flash File Systems is mounted successfully} 
	\While{SPIFFS.status() == FALSE}{ 
		SPIFFS.begin()
		\tcc{Open Certificate and Key files in flash memory of sensing node}
		SPIFFS.open("Root CA certificate($CA_{cert}$)") \;
		SPIFFS.open("Public Key (PuK)") \;
		SPIFFS.open("Private Key (PrK)") \;
		\tcc{Load certificate and key files}
		esp8266Client.loadCertificate(PuK) \;
		esp8266Client.loadPrivateKey(PrK) \;
		esp8266Client.loadCACert($CA_{cert}$) \;
	}
	\tcc{Attempting MQTT connection}
	\While{pubClient.connected() == FALSE}{ 
		\tcc{Trying to reconnect}
		pubClient.connect()\;
	}
	\tcc{Environmental and Geolocation inputs from sensors along with shipment overhead information is processed every 60 seconds}
	\For{every 60 seconds}{
		\tcc{JSON String is generetd including all environmental and geolocation data from sensing node}
		Publishing message MSG	$\leftarrow$ JSON(timestamp,lat,lng,sku,lot,drugName,temp,hum) \;
		\tcc{Data to be published is converted to char array}
		MSG+ $\leftarrow$ JSON.toCharArray(MSG)\;
		\tcc{Converted MSG+ is published to topic $\tau_{pub}$}
		pubclient.publish($\tau_{pub}$,MSG+)\;
	}
	\caption{Algorithm for Sensing Node in Proposed PharmaChain Implementation.}
	\label{ALG:Algorithm for Sensing Node in Proposed PharmChain Implementation}
\end{algorithm}

\begin{figure}[htbp]
	\centering
	\includegraphics[width=\textwidth]{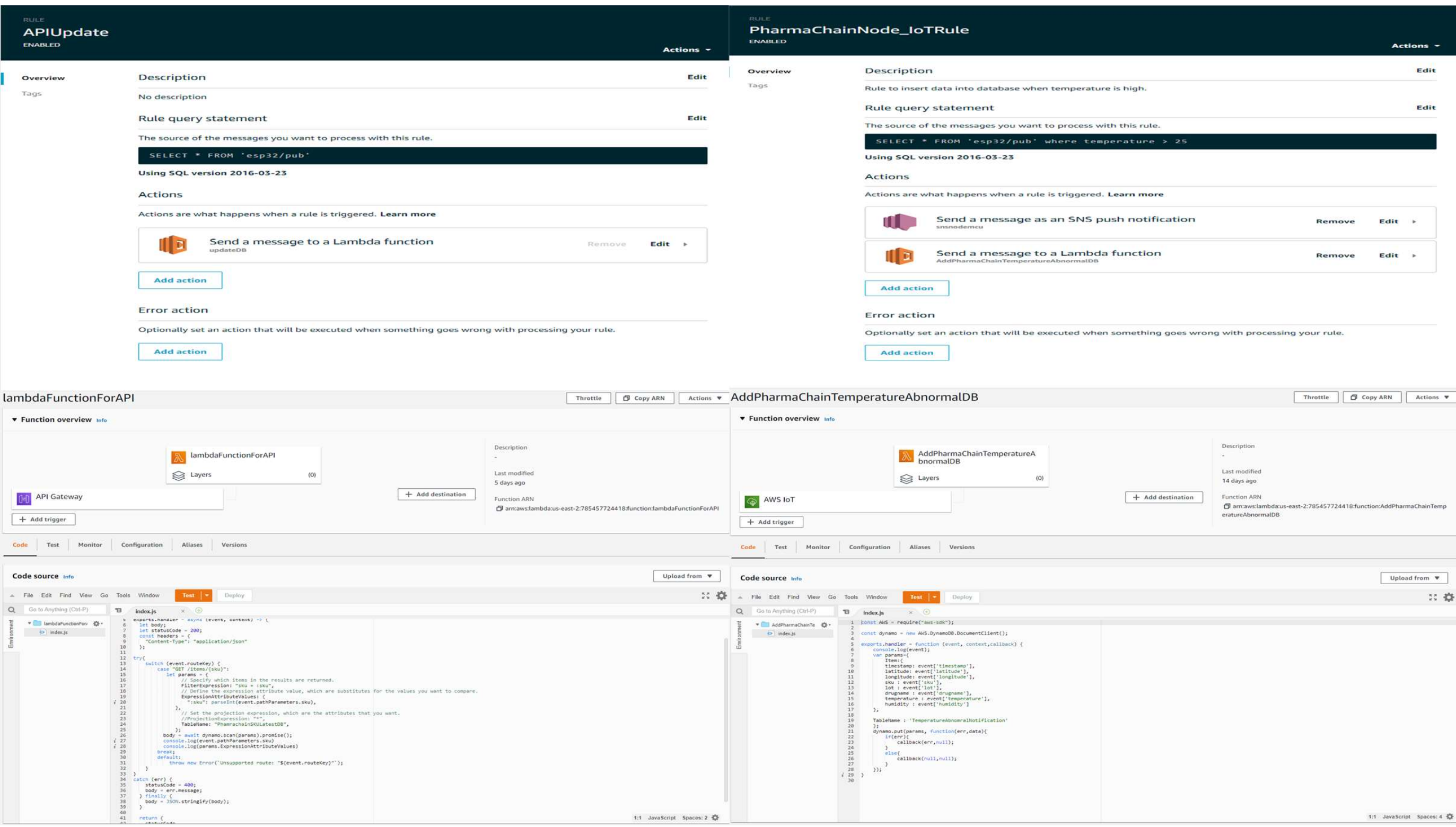}
	\caption{Implemented Data Source Rules and Functions in Proposed PharmaChain.}
	\label{FIG:Implemented Data Source Rules and Functions in Proposed PharmaChain}
\end{figure}

Data from the published topic is consumed by the cloud infrastructure which is going to process the received data based on a set of predefined rules and acts as data source for the oracle component. During the implementation of the prototype for an over the counter (OTC) available drug, for which storage temperature is less than 25$^{\circ}$C, is considered as an example. Once the data is consumed, it is checked for any temperature abnormalities (i.e., in this case if Temperature > 25$^{\circ}$C). If the temperature from the sensing node satisfies the rule, the data will be logged into a database for auditing and an immediate notification via e-mail is sent to the registered parties. The writing of data to the database and sending notifications is achieved by using cloud functions. Along with the abnormalities, the latest data is updated into another database table from which the data will be fetched when oracles make an API request. For fulfilling API requests, an API Gateway with required methods and path parameters is designed. As the data is only retrieved from the DB tables, an HTTP GET method, which will retrieve shipment details from the DB table based on SKU, is implemented. The steps for the implementation of the data source are shown in algorithm \ref{ALG:Algorithm for Data Source API Generation in Proposed PharmChain Implementation} and the implemented rules and functions can be seen in figure \ref{FIG:Implemented Data Source Rules and Functions in Proposed PharmaChain}.

\begin{algorithm}[!ht]
	\DontPrintSemicolon
	\KwInput{Input JSON message from Sensing Node (MSG+), HTTP GET API Request from Oracle}
	\KwOutput{Requested Shipment Data }
	\tcc{Data published MSG+ on topic $\tau_{pub}$ from previous algorithm is consumed by the IoT cloud component}
	\If{$\tau_{pub}$.hasMessage()}{
		MSG  $\leftarrow$ IoTCore.consume($\tau_{pub}$,MSG+) \;
		\tcc{In the example drug taken temperature threshold is 25$^{\circ}$C}
		\If{MSG.temperature > 25)}{
			\tcc{Update the audit table for temeprature abnormalities recorded during the supply chain process}
			AbnormalTempAuditTable.update(MSG)\;
			\tcc{An E-mail notitification will be sent to the registered parties in case the temperature exceeds the threshold}
			SNSNotification.sendEmail(MSG) \;
		}
	\tcc{Update this table with latest information of the shipment}
		ShipmentLatestUpdateTable.update(MSG)\;
	}
	\While{HTTP GET Request (req)}{
		\tcc{get request path parameters to geenrate the DB query}
		pathParameter param = req.getParam()\;
		\tcc{Scan the DB Table to retreive requested shipment information and store result}
		result = ShipmentLatestUpdateTable.scan(filtering param)\;
		\tcc{result from scan is mapped to http body}
		http.mapBody(result)\;
		\tcc{A http response is generated and sent back to the requested oracle}
		httpResponse (resp) $\leftarrow$ http.generateResponse(httpHeaders,httpBody) \;
		http.ResponseSend(resp)\;
	}
	\caption{Algorithm for Data Source API Generation in Proposed PharmChain Implementation.}
	\label{ALG:Algorithm for Data Source API Generation in Proposed PharmChain Implementation}
\end{algorithm}

The proposed blockchain architecture is implemented using Chainlink for oracle services and Ethereum as the blockchain platform to build the smart contracts. Chainlink is an oracle service which will enable Web 3.0 applications to access data from existing real-world data sources \cite{Chainlink_Whitepaper}. This will help in developing hybrid smart contracts which can interact with real-world systems and to provide real-world solutions. Ethereum is an open-source blockchain platform and also provides a Turing complete language (Solidity) to build decentralized applications. The programming environment provided by Ethereum works based on the Ethereum Virtual Machine (EVM) which is a world computer whose state is maintained and agreed upon by all the participant nodes in the P2P network. Smart contracts are executed within the EVM and these executions will move state of the EVM from one finite state to another. All the nodes in the network will record and store the state changes of the EVM to form a ledger of transactions, thus forming the blockchain. Ethereum currently works on a Proof-of-Work (PoW) consensus mechanism where the miner is chosen by posing a computationally hard hash problem to miner nodes. 

For building the prototype of the proposed PharmaChain, the KOVAN test network provided by Ethereum is used which works on Proof-of-Authority (PoA). Test networks are P2P networks which will simulate the main network of Ethereum in execution but utilize test tokens for all the payments and gas fees. Chainlink also provides different jobs in the KOVAN testnet which can be executed by utilizing a test LINK token instead of spending real money for deploying and testing decentralized applications. Smart contracts developed for PharmaChain can be logically divided into three components: smart contracts for implementing Role Based Access Control (RBAC), smart contracts to work with the Chainlink component and smart contracts for the core pharmaceutical supply chain mechanism. Smart contracts for RBAC will define different roles and will associate them with different Ethereum addresses. Responsibilities for each role are defined and controlled by defining modifiers in the Solidity language. Modifiers can also be defined to check some pre-defined conditions in the blockchain before executing the transactions submitted by a user. Before looking into defined modifiers, a shipment within the chain assumes different states which are enumerated as shown in Table \ref{TAB:Assumed Shipment States in Proposed PharmaChain and Corresponding Enumeration}. Below are the smart contract functions and modifiers defined for RBAC in the implemented PharmaChain prototype. 

\textbf{function addManufacturer(address account) public onlyManufacturer}: This function will add a new address to the list of manufacturers. It takes the address of the manufacturer as parameter and checks if the given address is not already associated with the role Manufacturer. If not associated, the address will be added to the list of manufacturers. 

\textbf{function renounceManufacturer() public}: This function helps in removing the access privileges of a manufacturer. The address from which the transaction is generated (msg.sender) will be checked for manufacturer privileges; if assigned, the role will be revoked and the address will be removed from list of manufacturers. 

\textbf{function isManufacturer(address account) public view returns (bool)}: This is a view function defined to check if the given address has the manufacturer role assigned. If assigned, it will return true, and false otherwise. 

\textbf{function addDistributor(address account) public onlyDistributor}: This function will add a new address to the list of Distributors. It takes the address of the distributor as parameter and checks if the given address is not already associated with the role distributor. If not associated, the address will be added to the list of distributors. 

\textbf{function renounceDistributor() public}: This function helps in removing the access privileges of the distributor. The address from which the transaction is generated (msg.sender) will be checked for distributor privileges; if assigned, the role will be revoked and the address will be removed from list of distributors. 

\textbf{function isDistributor(address account) public view returns (bool)}: This is a view function defined to check if the given address has a distributor role assigned. If assigned, it will return true, and false otherwise. 

\textbf{function addRetailer(address account) public onlyRetailer}: This function will add a new address to the list of retailers. It takes the address of the retailer as parameter and checks if the given address is not already associated with the role retailer. If not associated, the address will be added to the list of retailers. 

\textbf{function renounceRetailer() public}: This function helps in removing the access privileges of the retailer. The address from which the transaction is generated (msg.sender) will be checked for retailer privileges; if assigned, the role will be revoked and the address will be removed from list of retailers. 

\textbf{function isRetailer(address account) public view returns (bool)}: This is a view function defined to check if the given address has retailer role assigned. If assigned, it will return true, and false otherwise. 

\textbf{function addConsumer(address account) public onlyConsumer}: This function will add a new address to the list of consumers. It takes the address of the consumer as parameter and checks if the  given address is not already associated with role consumer. If not associated, the address will be added to the list of consumers. 

\textbf{function renounceConsumer() public}: This function helps in removing the access privileges of the consumer. The address from which the transaction is generated (msg.sender) will be checked for consumer privileges; if assigned the role will be revoked and the address will be removed from list of consumers. 

\textbf{function isConsumer(address account) public view returns (bool)}: This is a view function defined to check if the given address has consumer role assigned. If assigned, it will return true, and false otherwise. 

\textbf{modifier onlyOwner()}: This modifier is defined to check the ownership of the contract. This function checks if the sender of the transaction (msg.sender) is the same as the one who deployed this contract and can be used to control transfer of ownership of contract from one user to another.

\textbf{modifier verifyCaller(address \_address)}: This modifier is used to compare the sender of a transaction (msg.sender) to a given Ethereum address parameter and returns a boolean true if they match, and false otherwise. 

\textbf{modifier producedByManufacturer(uint \_upc)}:  This modifier will take the Universal Product Code (UPC) of a shipment and will check the state of the product in the supply chain. If the state of the product is '0', as shown in Table \ref{TAB:Assumed Shipment States in Proposed PharmaChain and Corresponding Enumeration}, it will return true and false otherwise. 

\textbf{modifier updateInventoryByManufacturer(uint \_upc)}: This modifier will take the UPC as parameter and checks if the state of the manufactured product is moved to Update Inventory By Manufacturer. If the state matches '1', it returns true and false otherwise. 

\textbf{modifier purchasedByDistributor (uint \_upc)}:  This modifier will take the UPC of a shipment and will check the state of the product in the supply chain. If the state of the product is '2', as shown in Table \ref{TAB:Assumed Shipment States in Proposed PharmaChain and Corresponding Enumeration}, it will return true and false otherwise.

\textbf{modifier shippedByManufacturer (uint \_upc)}:  This modifier will take the UPC of a shipment and will check the state of the product in the supply chain. If the state of the product is '3', as shown in Table \ref{TAB:Assumed Shipment States in Proposed PharmaChain and Corresponding Enumeration}, it will return true and false otherwise.

\textbf{modifier receivedByDistributor (uint \_upc)}:  This modifier will take the UPC of a shipment and will check the state of the product in the supply chain. If the state of the product is '4', as shown in Table \ref{TAB:Assumed Shipment States in Proposed PharmaChain and Corresponding Enumeration}, it will return true and false otherwise.

\textbf{modifier processByDistributor (uint \_upc)}:  This modifier will take the UPC of a shipment and will check the state of the product in the supply chain. If the state of the product is '5', as shown in Table \ref{TAB:Assumed Shipment States in Proposed PharmaChain and Corresponding Enumeration}, it will return true and false otherwise.

\textbf{modifier packagedByDistributor (uint \_upc)}:  This modifier will take the UPC of a shipment and will check the state of the product in the supply chain. If the state of the product is '6', as shown in Table \ref{TAB:Assumed Shipment States in Proposed PharmaChain and Corresponding Enumeration}, it will return true and false otherwise.

\textbf{modifier forSaleByDistributor (uint \_upc)}:  This modifier will take the UPC of a shipment and will check the state of the product in the supply chain. If the state of the product is '7', as shown in Table \ref{TAB:Assumed Shipment States in Proposed PharmaChain and Corresponding Enumeration}, it will return true and false otherwise.

\textbf{modifier shippedByDistributor (uint \_upc)}:  This modifier will take the UPC of a shipment and will check the state of the product in the supply chain. If the state of the product is '8', as shown in Table \ref{TAB:Assumed Shipment States in Proposed PharmaChain and Corresponding Enumeration}, it will return true and false otherwise.

\textbf{modifier purchasedByRetailer (uint \_upc)}:  This modifier will take the UPC of a shipment and will check the state of the product in the supply chain. If the state of the product is '9', as shown in Table \ref{TAB:Assumed Shipment States in Proposed PharmaChain and Corresponding Enumeration}, it will return true and false otherwise.

\textbf{modifier receivedByRetailer (uint \_upc)}:  This modifier will take the UPC of a shipment and will check the state of the product in the supply chain. If the state of the product is '10', as shown in Table \ref{TAB:Assumed Shipment States in Proposed PharmaChain and Corresponding Enumeration}, it will return true and false otherwise.

\textbf{modifier forSaleByRetailer (uint \_upc)}:  This modifier will take the UPC of a shipment and will check the state of the product in the supply chain. If the state of the product is '11', as shown in Table \ref{TAB:Assumed Shipment States in Proposed PharmaChain and Corresponding Enumeration}, it will return true and false otherwise.

\textbf{modifier purchasedByConsumer (uint \_upc)}:  This modifier will take the UPC of a shipment and will check the state of the product in the supply chain. If the state of the product is '12', as shown in Table \ref{TAB:Assumed Shipment States in Proposed PharmaChain and Corresponding Enumeration}, it will return true and false otherwise.

\textbf{modifier onlyManufacturer()}: This modifier is used to check if the sender of a transaction is associated with the role manufacturer. The function isManufacturer(address account) is called with address parameter  msg.sender and a boolean result true is returned if the role is assigned or false otherwise. 

\textbf{modifier onlyDistributor()}: Similar to the previous function isDistributor(address account) is called with address parameter from msg.sender and true is returned if the role is assigned or false otherwise. 

\textbf{modifier onlyRetailer()}: This modifier is used to check if the sender of a transaction is associated with the role retailer. The function isRetailer(address account) is called with address parameter msg.sender and a boolean result true is returned if the role is assigned or false otherwise.

\textbf{modifier onlyConsumer()}: This modifier is used to check if the sender of transaction is associated with the role consumer. The function isConsumer(address account) is called with address parameter msg.sender and a boolean result true is returned if the role is assigned or false otherwise.

\begin{table}[htbp]
	\caption{Assumed Shipment States in Proposed PharmaChain and Corresponding Enumeration.}
	\label{TAB:Assumed Shipment States in Proposed PharmaChain and Corresponding Enumeration}
	\centering
	\begin{tabular}{|p{0.3\textwidth}|p{0.45\textwidth}|p{0.15\textwidth}|}
		\hline
		\textbf{Pharmaceutical Shipment State} & \textbf{Description} & \textbf{Enumerated State Value} \\
		\hline
		Produced By Manufacturer & Pharmaceutical Manufacturer acquires necessary RAW materials and drugs are manufactured & 0 \\
		\hline
		Update Inventory By Manufacturer & Inventory is updated by manufactured and offered for sale & 1\\
		\hline
		 Purchased By Distributor& Pharmaceutical Distributor purchases from inventory of manufacturer based on market demand & 2\\
		\hline
		 Shipped By Manufacturer & After successful purchase product is shipped from manufacturer to distributor & 3\\
		\hline
		 Received By Distributor & Purchased product is received by Distributor &4\\
		\hline
		 Processed By Distributor & Inventory update at distributor &5\\
		\hline
	 	 Packaged By Distributor & Processed product is sometimes re-packaged or branded by distributor & 6\\
		\hline
		For Sale By Distributor & Inventory is set up for sale by the distributor & 7\\
		\hline
		Purchased By Retailer & Product is purchased by In-mail/ Brick and Mortar retailer pharmacies & 8\\
		\hline
		Shipped By Distributor & Once purchase is successful, product is moved from distributor to retailer locations &9\\
		\hline
		Received By Retailer & Product is received by retailer pharmacies &10\\
		\hline
		For Sale By Retailer & Inventory update and offer it for sale either In-mail/physical pharmacy & 11\\
		\hline
		Purchased By Consumer & Consumer purchases required medication from phamracies &12\\
		\hline
	\end{tabular}
\end{table}

Functions implemented for pharmaceutical supply chain are discussed below and their associated access control modifiers can be seen in Table \ref{TAB:PharmaChain Smart Contract Functions and Associated Modifiers and Event Generations}. 

\textbf{requestTemperatureData(string memory \_sku)}: This function helps to interact with the Chainlink oracle to retrieve the current ambient temperature around the shipment from the sensing node. It takes a SKU as a parameter and this value will be send to  API Get request as path parameter, based on which the cloud DB tables will be queried to get the latest temperature value. This function can be called by any account with any role or no role attached.

\textbf{requestHumidityData(string memory \_sku)}: This function helps to interact with the Chainlink oracle to retrieve the current relative humidity around the shipment from the sensing node. It takes a SKU as a parameter and this value will be send to  API Get request as path parameter, based on which the cloud DB tables will be queried to get the latest temperature value. This function can be called by any account with any role or no role attached.

\textbf{requestLatitude(string memory \_sku)}: This function helps to interact with the Chainlink oracle to retrieve the current latitude of the shipment from the sensing node. It takes a SKU as a parameter and this value will be send to  API Get request as path parameter, based on which the cloud DB tables will be queried to get the latest temperature value. This function can be called by any account with any role or no role attached.

\textbf{requestLongitude(string memory \_sku)}: This function helps to interact with the Chainlink oracle to retrieve the current longitude of the shipment from the sensing node. It takes a SKU as a parameter and this value will be send to  API Get request as path parameter, based on which the cloud DB tables will be queried to get the latest temperature value. This function can be called by any account with any role or no role attached.

\textbf{verifyAuthenticityoFProduct(uint \_upc)}: This function will verify the origin of the product along with the product transfer between manufacturer, distributor, retailer and consumer to prove the authenticity of the product. Authenticity can be verified at any stage of the supply chain, hence can be called from any account with or without any role attached. 

\textbf{produceItemByManufacturer(string memory \_sku, string memory \_drugName, uint \_upc)}: This function is executed by a manufacturer for each newly produced product. Each product with assigned UPC and SKU along with drug name will be sent as parameters to this function. This will create a new product in the PSC. This function can only be executed by accounts with role manufacturer and generates an event "ProducedByManufacturer" to be recorded in the blockchain. 

\textbf{sellItemByManufacturer(uint \_upc)}: This function is executed by the manufacturer after producing new products. Before executing this function, it will check if the caller of the function is  manufacturer and really owns the product to sell. Once verified it will change the state of the product to available for selling and generates an event to record in the blockchain. 

\textbf{purchaseItemByDistributor(uint \_upc)}: This function helps distributors to buy products offered for sale by manufacturers. Any account with distributor role can call this function and once the purchase is done, the status of the product along with the ownership of the product will change from manufacturer to purchased distributor.

\textbf{shippedItemByManufacturer(uint \_upc)}: After successful purchase, the manufacturer is responsible for shipping the product and this function is used to update the status of product to shipped and can be called by the manufacturer who is assigned as origin manufacturer of the product. This generates an event shipped to be recorded in the blockchain. 

\textbf{receivedItemByDistributor(uint \_upc)}: This function is used by a distributor to update the shipment once received from manufacturer. This function verifies if the caller is a distributor and actually purchased the product before updating the status of shipment to ``received''. This function will generate a received shipment event to be recorded in the blockchain.  

\textbf{processedItemByDistributor(uint \_upc)}: Once shipment is received by a distributor and processed, this function is used to update the status of the shipment to processed. This function can only be executed by the distributor who owns the shipment. 

\textbf{packageItemByDistributor(uint \_upc)}: Sometimes distributors re-package the product before offering to sell it. Once repackaging is done this function is called by the distributor to change the state of the product to be re-packaged and ready to be sold. This function ensures the distributor who actually owns the product can make the function call. 

\textbf{sellItemByDistributor(uint \_upc)}: This function helps a distributor to offer the product owned for sale. Only the distributor who owns the product can make a call to this function and if the status of the product is already packaged by distributor. A sell event is generated by this function to be recorded in the blockchain. 

\textbf{purchaseItemByRetailer(uint \_upc)}: This function helps any user with retailer role assigned to make a purchase from a distributor. Once purchase is done, the owner of the product will be changed to the purchased retailer and an event is generated to be recorded in the blockchain. 

\textbf{shippedItemByDistributor(uint \_upc)}: After the purchase by a retailer, the shipment is shipped from distributor to the retailer. This event is updated by using this function and a ship event is generated to be recorded in the blockchain. 

\textbf{receivedItemByRetailer(uint \_upc)}: Once the shipment is received from the distributor, the retailer will execute this function to record the shipment received event in the blockchain. 

\textbf{sellItemByRetailer(uint \_upc)}: Once a retailer receives the shipment and processes it, it is offered for sale by using this function. This function, before executing, will ensure that the caller of the function is the retailer who actually owns the product. An event is generated to be recorded in the blockchain. 

\textbf{purchaseItemByConsumer(uint \_upc)}: This is the final step in the supply chain where the consumer can make the purchase of the product from a retailer. Any user with consumer role attached can execute this function which will change the ownership of the product from retailer to consumer along with generating a purchase event to be recorded in blockchain.

\begin{table}[htbp]
	\caption{PharmaChain Smart Contract Functions and Associated Access Control Modifiers and Event Generations.}
	\label{TAB:PharmaChain Smart Contract Functions and Associated Modifiers and Event Generations}
	\centering
	\begin{tabular}{|p{0.3\textwidth}|p{0.3\textwidth}|p{0.30\textwidth}|}
		\hline
		\textbf{Pharmaceutical Supply Chain Smart \linebreak Contract Function} & \textbf{Modifiers} & \textbf{Event Generated} \\
\hline \hline
		requestTemperatureData(string memory \_sku) & Any & None \\
		\hline
		requestHumidityData(string memory \_sku) & Any & None\\
		\hline
		requestLatitude(string memory \_sku) & Any & None\\
		\hline
		requestLongitude(string memory \_sku) & Any & None\\
		\hline
		verifyAuthenticityoFProduct(uint \_upc) & Any & None\\
		\hline
		produceItemByManufacturer(string memory \_sku, string memory \_drugName, uint \_upc) & onlyManufacturer() & ProducedByManufacturer(uint upc)\\
		\hline
		sellItemByManufacturer(uint \_upc) & onlyManufacturer() \linebreak producedByManufacturer(\_upc) \linebreak verifyCaller(items[\_upc].ownerID) & UpdateInventoryByManufacturer(uint upc)\\
		\hline
		purchaseItemByDistributor(uint \_upc) & onlyDistributor() \linebreak updateInventoryByManufacturer(\_upc) & PurchasedByDistributor(uint upc)\\
		\hline
		shippedItemByManufacturer(uint \_upc) & onlyManufacturer() \linebreak purchasedByDistributor(\_upc) \linebreak verifyCaller(items[\_upc].originManufacturerID) & ShippedByManufacturer(uint upc)\\
		\hline\
		receivedItemByDistributor(uint \_upc) & onlyDistributor() \linebreak shippedByManufacturer(\_upc) \linebreak verifyCaller(items[\_upc].ownerID) & ReceivedByDistributor(uint upc)\\
		\hline
		processedItemByDistributor(uint \_upc) & onlyDistributor() \linebreak receivedByDistributor(\_upc) \linebreak verifyCaller(items[\_upc].ownerID) & ProcessedByDistributor(uint upc)\\
		\hline
		packageItemByDistributor(uint \_upc) & onlyDistributor() \linebreak processByDistributor(\_upc) \linebreak verifyCaller(items[\_upc].ownerID) & PackagedByDistributor(uint upc)\\
		\hline
		sellItemByDistributor(uint \_upc) & onlyDistributor() \linebreak packagedByDistributor(\_upc) \linebreak verifyCaller(items[\_upc].ownerID)& ForSaleByDistributor(uint upc)\\
		\hline
		purchaseItemByRetailer(uint \_upc) & onlyRetailer() \linebreak forSaleByDistributor(\_upc) & PurchasedByRetailer(uint upc)\\
		\hline
		shippedItemByDistributor(uint \_upc) & onlyDistributor() \linebreak purchasedByRetailer(\_upc) \linebreak verifyCaller(items[\_upc].distributorID) & ShippedByDistributor(uint upc)\\
		\hline
		receivedItemByRetailer(uint \_upc) & onlyRetailer() \linebreak shippedByDistributor(\_upc) \linebreak verifyCaller(items[\_upc].ownerID) & ReceivedByRetailer(uint upc)\\
		\hline
		sellItemByRetailer(uint \_upc) & onlyRetailer() \linebreak receivedByRetailer(\_upc) \linebreak verifyCaller(items[\_upc].ownerID) & ForSaleByRetailer(uint upc)\\
		\hline
		purchaseItemByConsumer(uint \_upc) & onlyConsumer() & PurchasedByConsumer(uint upc)\\
		\hline
	\end{tabular}
\end{table}

\begin{figure}[htbp]
	\centering
	\includegraphics[width=\textwidth]{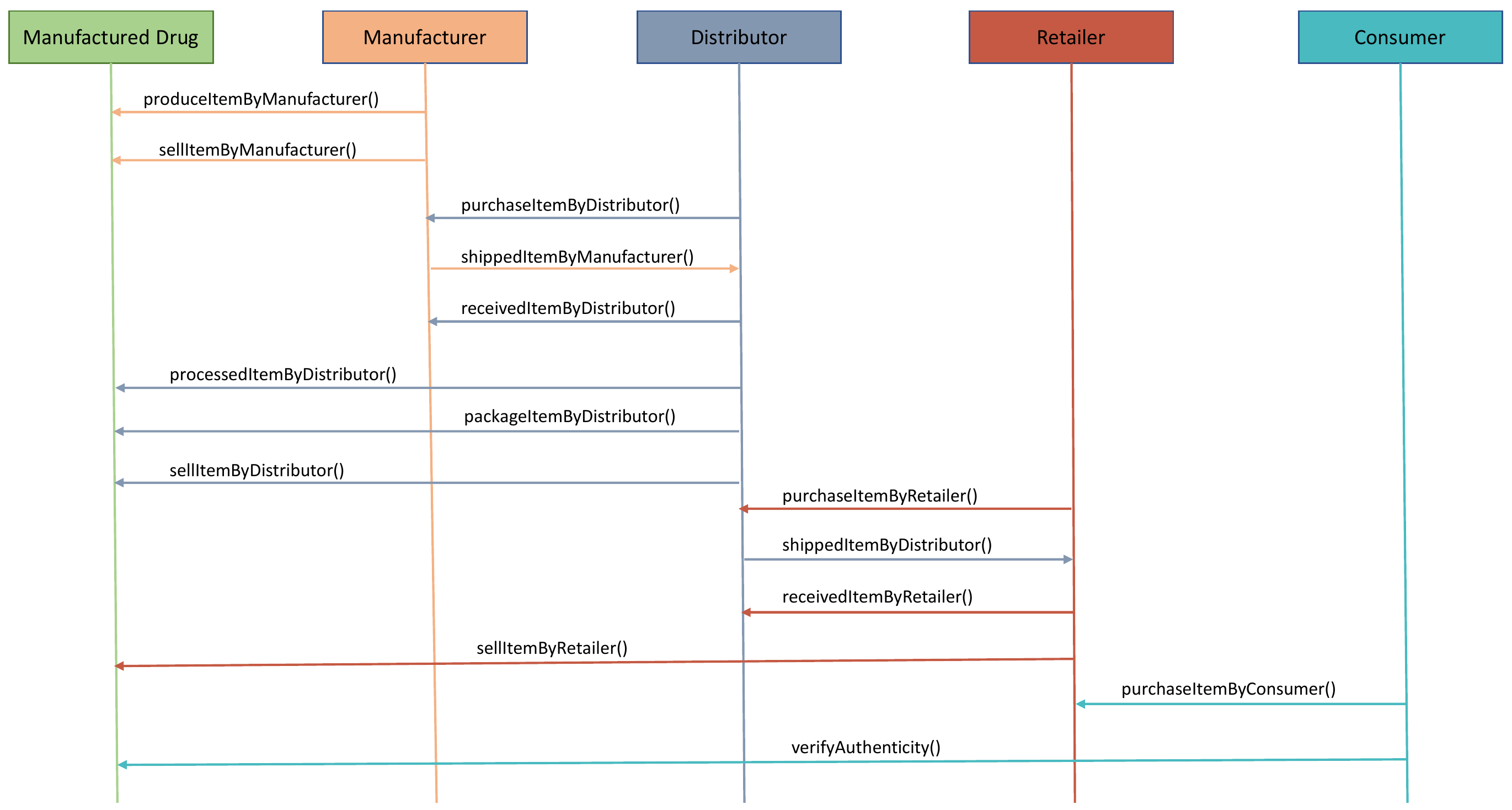}
\caption{Sequence Diagram of Implemented PharmaChain Showing Interactions Between Entities Using Smart Contract Functions.}
	\label{FIG:Sequence Diagram of Implemented PharmaChain Showing Interactions Between Entities Using Smart Contract Functions}
\end{figure}

The implemented PharmaChain application sequence of execution steps is shown in figure \ref{FIG:Sequence Diagram of Implemented PharmaChain Showing Interactions Between Entities Using Smart Contract Functions}. During the first step the manufacturer creates a new product which has its own SKU and UPC assigned by the manufacturer. Once the product is available, the manufacturer will update the inventory and will offer the product for sale. This will make the product available for purchase by any distributor. A distributor purchases the product which will change the ownership of the product from manufacturer to the distributor. The manufacturer then ships the product to the purchasing distributor and updates the status. Once the distributor receives the product, processes and packages the product before selling it to the retailer. The retailer will then make the purchase and the ownership of product will change from the distributor to retailer. Once the retailer receives the product, will offer it for sale to the customer. In the final step, consumer will be able to purchase it and also check the authenticity of the product by verifying the ownership changes between entities.

\begin{table}[htbp]
	\caption{Deployed PharmaChain Smart Contract Details.}
	\label{TAB:Deployed PharmaChain Smart Contract Details}
	\centering
	\begin{tabular}{|p{0.2\textwidth}|p{0.7\textwidth}|}
		\hline
\textbf{Parameters} & \textbf{Value} \\
\hline		\hline
		Deployed Contract Address & 0x9409739e0D68Cb459a41892a48F3E62A2ceb7eeF \\
		\hline
		Contract Owner Address & 0xBc6251e4f6389117a8f08b9B465a0bd44bA15Ab4\\
		\hline
		Manufacturer Address & 0x3eDe97Ea0DFF3EcD7320b1822E33f4a2764E8ed4 \\
		\hline
		Distributor Address & 0xE8FDEa8272393Ad05f004d2BF583D784b509f9D3 \\
		\hline
		Retailer Address & 0x81F66b24db9EA43f454636750922c4ef12c26b48 \\
		\hline
		Consumer Address & 0xD46c558431c9CA642A85885509FBf6dA4Ba60435 \\
		\hline 
		Contract Creation Transaction Hash & 0x0c99850a32d2f836023a8ada0ec3feab2ae34be2d16969c2d38537e51e77db7a \\
		\hline
		Oracle Address for Unsigned Integer data & 0xc57B33452b4F7BB189bB5AfaE9cc4aBa1f7a4FD8 \\
		\hline
		Job ID for Unsigned Integer data & d5270d1c311941d0b08bead21fea7747 \\
		\hline
		Oracle Address for Signed Integer data & 0x9C0383DE842A3A0f403b0021F6F85756524d5599 \\
		\hline
		Job ID for Signed Integer data & ba1d5d5070a247eaa7070f838a42bb03 \\
		\hline
	\end{tabular}
\end{table}

\begin{figure}[t]
	\centering
	\includegraphics[width=\textwidth]{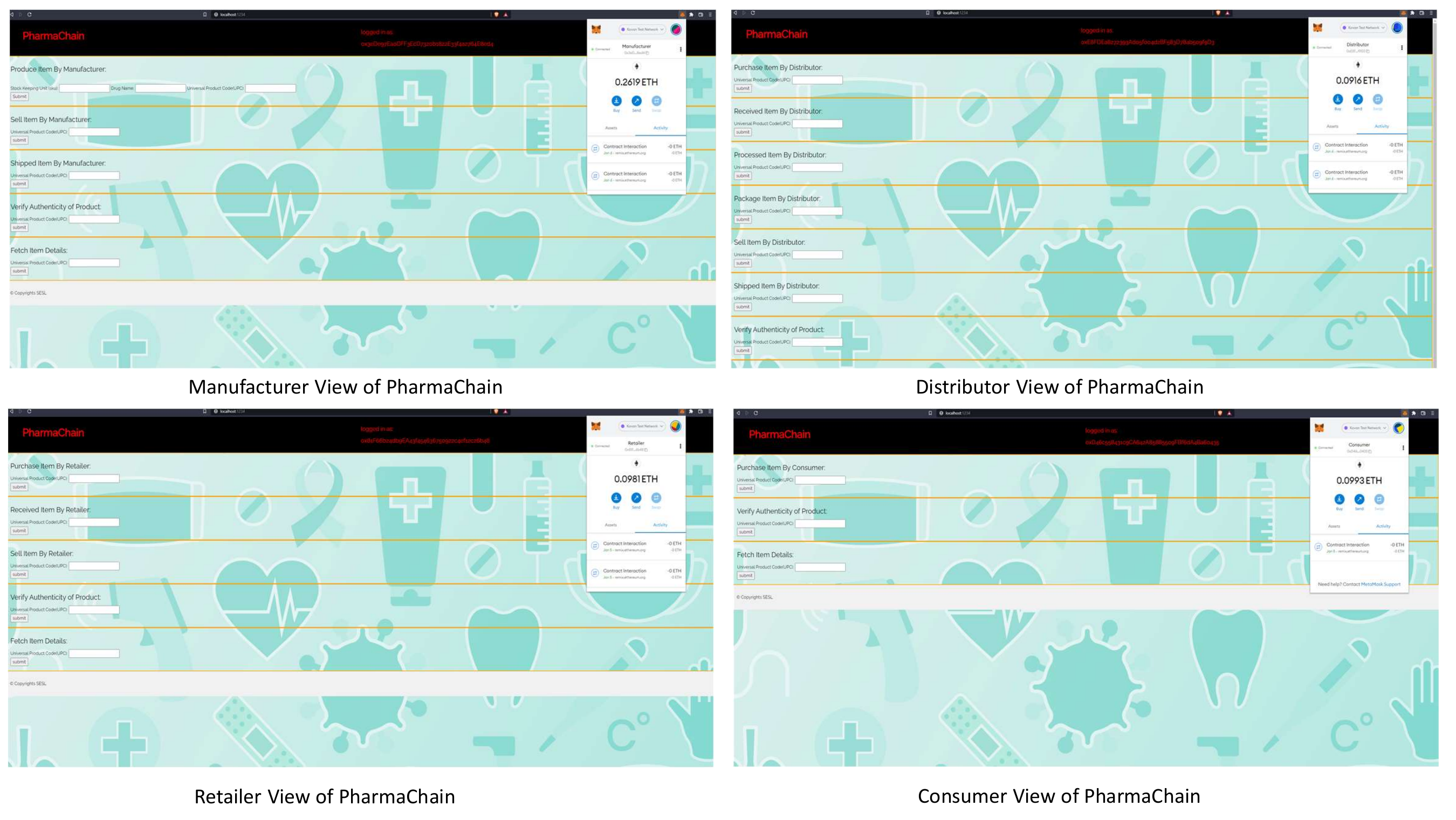}
	\caption{User Interface for Implemented PharmaChain based on Different Roles.}
	\label{FIG:User Interface for Implemented PharmaChain based on Different Roles}
\end{figure}

The User Interface (UI) is implemented using the web3 library to interact with published smart contracts from the KOVAN test network. Metamask is used as wallet provider to operate and utilize test currency for deploying and testing the functionality of the application. Based on roles, different entities have access to different functions within the supply chain, as shown in figure \ref{FIG:User Interface for Implemented PharmaChain based on Different Roles}. The manufacturer has only access to ``Produce Item By Manufacturer'', ``Sell Item By Manufacturer'', and ``Ship Item By Manufacturer''. Similarly a distributor has access to ``Purchase Item By Distributor'', ``Received Item By Distributor'', ``Processed Item By Distributor'', ``Package Item By Distributor'', ``Sell Item By Distributor'' and ``Shipped Item By Distributor''. A retailer has access to ``Purchase Item By Retailer'', ``Received Item By Retailer'', and ``Sell Item By Retailer''. A consumer has access to ``Purchase Item By Consumer''. All the roles have access to ``Fetch Item Details'' and ``Verify Authenticity of Product''. Different transactions and transaction costs are evaluated in Table \ref{TAB:Performance Evaluation of Implemented PharmaChain}. These transaction costs can be avoided by implementing a private blockchain and private nodes for Chainlink.

\begin{table}[t]
	\caption{Performance Evaluation of Implemented PharmaChain.}
	\label{TAB:Performance Evaluation of Implemented PharmaChain}
	\centering
	\begin{tabular}{|p{0.15\textwidth}|p{0.12\textwidth}|p{0.4\textwidth}|p{0.1\textwidth}|p{0.08\textwidth}|}
		\hline
\textbf{Action} & \textbf{Tx fees for Ethereum Blockchain} & \textbf{Tx hash} & \textbf{Tx fees for Chainlink Oracle} & \textbf{Block Time (s)} \\ 
\hline \hline
		Contract Deployment & 0.00898 ETH & 0x0c99850a32d2f836023a8ada0ec3feab2ae34be\linebreak2d16969c2d38537e51e77db7a & 0 & 8\\
		\hline
		Add Manufacturer & 0.00011 ETH & 0xeae4963ac8b822d35f18bba2c20d5d01ee3ad4e\linebreak b2451c6372c3d3f93e7b11b47 & 0 & 4 \\
		\hline
		Add Distributor & 0.00011 ETH & 0x23c223c177185e78e4e7ab90b8bcf217930ebc5\linebreak7ebdcbce3431bfa366cd51bf9 & 0 & 4 \\
		\hline
		Add Retailer & 0.00011 ETH & 0x28aa318991ea876d5bd2bd7238577068adc690\linebreak131af94ee06d9e61b01e41391d & 0 & 8 \\
		\hline
		Add Consumer & 0.00011 ETH & 0xd6d91da84aa9f4d9346a7ab7ea088816dbde750\linebreak d566e8eeeb7f23bdf4489b442 & 0 & 4 \\
		\hline 
		Produce Item By\linebreak Manufacturer & 0.00151 ETH & 0x28169fde8509c6f2996763eba72280b2b15133\linebreak7d74143b5579e91a3ca8655aa7 & 0.5 LINK & 4 \\
		\hline
		Sell Item By\linebreak Manufacturer & 0.00143 ETH & 0x79048f114ea5c0ed737753f973f69c575a4cdbd\linebreak206d2cd33c69db97c5098f19a & 0.5 LINK & 8 \\
		\hline
		Purchase Item By\linebreak Distributor & 0.00118 ETH & 0xeab6ade50d16201b681a95a645499b0ae189f2d\linebreak4738c670ab081cb2a39f34d42 & 0.4 LINK & 4 \\
		\hline
		Shipped Item By\linebreak Manufacturer & 0.00106 ETH & 0xa52f4a9629c69c40c6550c00f903d1e63133859\linebreak6184e4552fa27841f8a9860a3  & 0.4 LINK & 4 \\
		\hline
		Received Item By\linebreak Distributor & 0.00106 ETH & 0x97eeb340ff9fb6b87cb006f4d4005043d7a809f\linebreak0479de85ad84024ace520ce43  & 0.4 LINK& 8\\
		\hline 
		Processed Item By\linebreak Distributor & 0.00106 ETH & 0x6ef2229a3947e4dde0dbe4f800158f78ff1fe9ed\linebreak e1300274f5c74b537218165a  & 0.4 LINK& 4\\
		\hline
		Packaged Item By\linebreak Distributor & 0.00106 ETH & 0xb11966bd5087d70c0baa6ead1e76e6b331b9d7\linebreak08aa43b8d8d82a997a2c185c3b & 0.4 LINK & 8 \\
		\hline
		Sell Item By\linebreak Distributor & 0.00107 ETH & 0xf7c0c2cac67f765a4fadaac4f935bd088b907d1c\linebreak ff4be0008328c0c1a68b6918 & 0.4 LINK & 4 \\
		\hline 
		Purchase Item By\linebreak Retailer & 0.00118 ETH & 0x25961c30875f3d9387c27dcf38b570fbccfd7cdf\linebreak c371404dc2f226c3d2d7d2fc & 0.4 LINK & 4 \\
		\hline 
		Ship Item By\linebreak Distributor & 0.00106 ETH & 0xb6d6fc63e6d5d21e53468f0d91193175f623c66\linebreak d649e1f4df028d5769c0c536e & 0.4 LINK & 4 \\
		\hline 
		Receive Item By\linebreak Retailer & 0.00106 ETH & 0xd122e0c86688659a386a83a0a87c34aebf6db1a\linebreak8a820676ef14413ae03b886cb & 0.4 LINK & 8\\
		\hline
		Sell Item By\linebreak Retailer & 0.00106 ETH & 0x13f1f0c73363b685445e54956d0f5667a7c0987\linebreak aa1290bd1f26ef342d68ba275  & 0.4 LINK & 4\\
		\hline
		Purchase Item By\linebreak Consumer & 0.00118 ETH & 0x3e522790ae54650b3880f5bfcbdf9c437f44b12\linebreak a8298147fc34f8de417460094 & 0.4 LINK & 4 \\
		\hline 
	\end{tabular}
\end{table}

\begin{figure}[htbp]
	\centering
	\includegraphics[width=\textwidth]{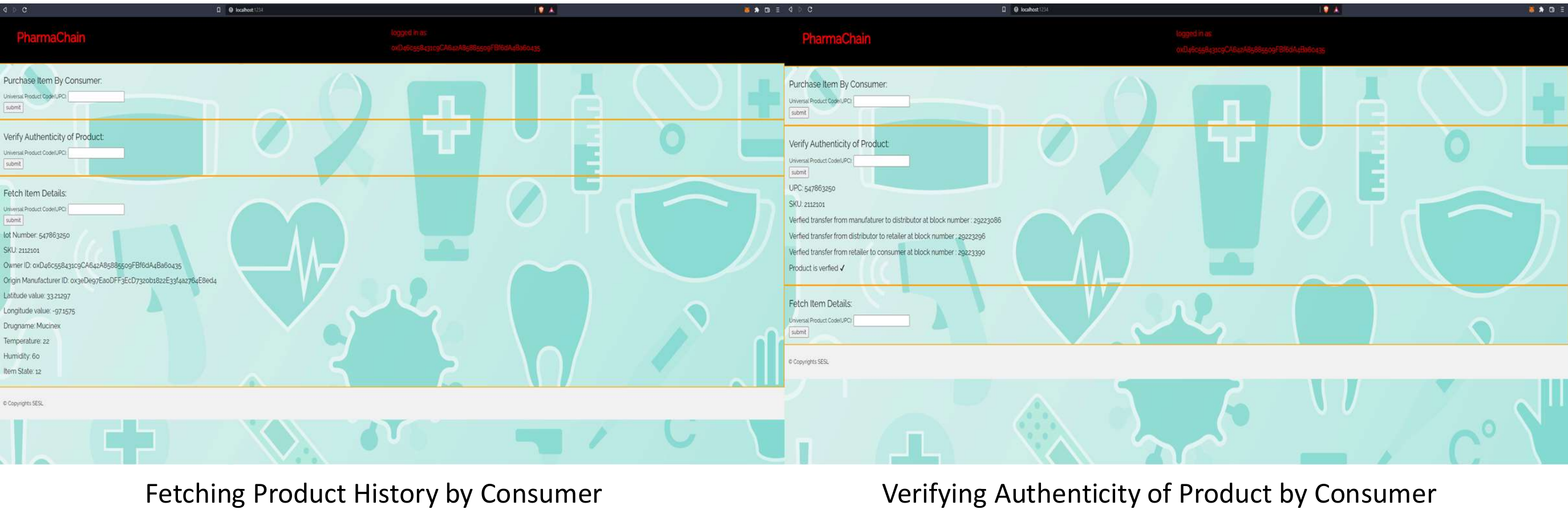}
	\caption{Consumer Verifying Product Details and Authenticity in Proposed PharmaChain.}
	\label{FIG:Consumer Verifying Product Details and Authenticity in Proposed PharmaChain}
\end{figure}

\subsection{Validation of PharmaChain}\label{Validation of PharmaChain}

\begin{figure}[htbp]
	\centering
	\includegraphics[width=\textwidth]{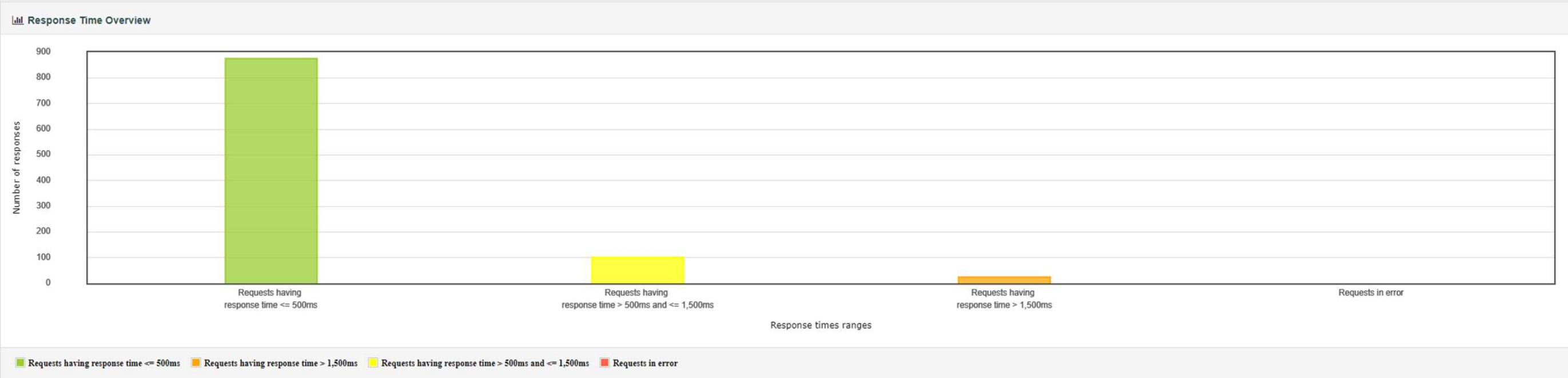}
	\caption{Response Time Overview of Load Test Performed on Data Cource Implemented in PharmaChain.}
	\label{FIG:Response Time Overview of Load Test Performed on Data Cource Implemented in PharmaChain}
\end{figure}

Transaction time is major factor which needs to be evaluated for scalability of the proposed PharmaChain system. As  can be seen in Table \ref{TAB:Performance Evaluation of Implemented PharmaChain}, the average transaction time is 5.6 sec. in the implemented prototype. Transaction times mainly depend on two components: mining times and Chainlink interactions, as the data has to be fetched from the given API. Mining times mainly depend on the transaction volume in the blockchain at that time, but it can be significantly reduced by adopting a private blockchain dedicated to process only transactions from the PSC. The prototype is implemented on a shared test network hence, the performance of the proposed application mainly depends on the performance of the API, as a blockchain test network cannot be controlled or fine tuned for performance. A load test is performed on the data source to determine the scalability and adaptability of the proposed system. To perform load tests, JMeter is used. A load test plan with 100 threads and loop count of 10 is used, which means that 1000 requests are sent to the cloud data source within a span of 2 seconds. From the results in Table \ref{TAB:Statistics of Load Testing Performed on Data Source Implemented in Proposed PharmaChain} it can be seen that the number of failed transactions is 0 with an average response time of 285ms, which is very low  considering that a large number of oracle requests are sent within a short span. The graph in Figure \ref{FIG:Response Time Overview of Load Test Performed on Data Cource Implemented in PharmaChain} shows that a large portion of the requests are successful within 500ms, making the proposed architecture for data source scalable and adaptable in real-world PSC.

\begin{table}[htbp]
	\caption{Statistics of Load Testing Performed on Data Source Implemented in Proposed PharmaChain.}
	\label{TAB:Statistics of Load Testing Performed on Data Source Implemented in Proposed PharmaChain}
	\centering
	\begin{tabular}{|p{0.5\textwidth}|p{0.35\textwidth}|}
		\hline
		\textbf{Parameters} & \textbf{Value}\\
\hline		\hline
		\textbf{Number of Oracle Requests sent} & 1000 \\
		\hline
		\textbf{Load Duration} & 2 Seconds \\
		\hline
		\textbf{Failed Requests} & 0 \\
		\hline
		\textbf{Percentage of Error} & 0\%\\
		\hline
		\textbf{Average Response Time(ms)} & 285.196 ms\\
		\hline
		\textbf{Minimum Response Time(ms) } & 78 ms\\
		\hline
		\textbf{Maximum Response Time(ms)} & 1960 ms\\
		\hline
		\textbf{Throughput (Requests/Sec)} & 16.66\\
		\hline
	\end{tabular}
\end{table}		

\section{Conclusion and Future Research} 
\label{Conclusion}

Along with providing a transparent pharmaceutical supply chain, which will enhance the processing times of  shipments and smooth operation between different distributed entities, the proposed PharmaChain also equips consumers with very much needed tools to verify the authenticity of the purchased drugs before consumption. By building a transparent chain, the proposed PharmaChain is also enhancing the accountability of the entities involved in the supply chain, thereby making it easy to trace and take severe actions against  entities with malicious intent. IoT technology was seamlessly integrated with the blockchain in the proposed system using oracles, along with real-time alerts to the stakeholders regarding different abnormal levels of environmental parameters like ambient temperature and relative humidity around the shipment. This helps in taking prompt actions to control the environment around the package and to ensure the safe handling of the product throughout the supply chain. A robust Role Based Access Control Mechanism was also implemented in the proposed PharmaChain using smart contracts. A proof-of-concept for the whole proposed system has been implemented and analyzed for scalability and reliability. The proof-of-concept was implemented using an Ethereum blockchain with Proof-of-Authority (PoA) as consensus mechanism and analysis of smart contract interactions has shown the average block time is 5.6sec which is acceptable for applications in the supply chain. To address the problem of smart contracts interacting with external APIs, oracles were used. A hybrid smart contract was built in PharmaChain which is capable of interacting with external IoT APIs and of consuming data into smart contracts which ensures the reliability of data being consumed in smart contracts. Scalability of the data source for oracles was also analyzed and the average response time is measured to be 285.196 ms hence proving the proposed PharmaChain application is reliable and scalable to be adopted as a solution to avoiding counterfeit medications in the PSU.

The current work can be extended further by introducing hardware security mechanisms into place like Physical Unclonable Functions (PUF) \cite{Yanambaka2018} instead of RSA encryption, which will increase the security of the IoT systems integrated with blockchain and will also reduce power consumption thereby reducing the cost of operations as there will be a very large number of sensing nodes needed throughout the supply chain. We have proposed the idea of PUFchain as an initial contribution in this regard \cite{8977825}. Most of the supply chain actions are automated but still few functions are not automated in the  current implementation. An IoT mechanism will be implemented in future work to automate the entire handling of medical drugs through the supply chain. In this direction, we think integration of pharmaceutical supply chain in smart healthcare \cite{8726143} as well as smart cities \cite{9194280} will have great impact. Thus, research is necessary in these various directions.

\bibliographystyle{IEEEtran}


\begin{IEEEbiography}
[{\includegraphics[height=1.3in,keepaspectratio]{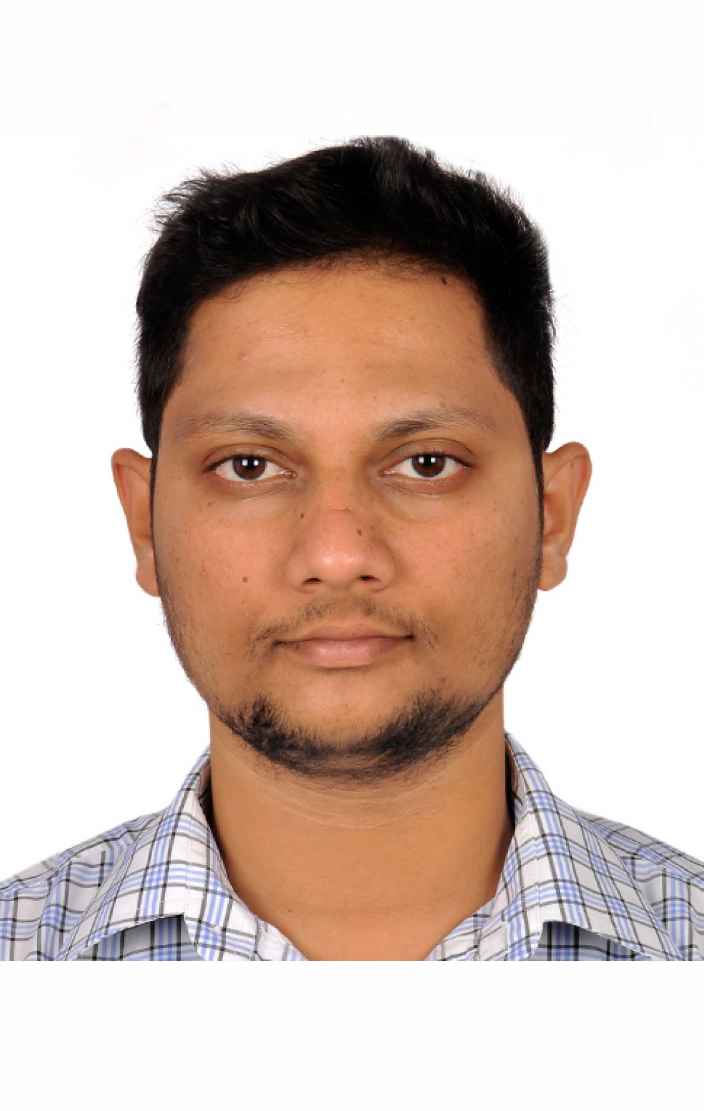}}] 
{Anand K. Bapatla} received a Bachelor's of Technology (B. Tech) in Electronics and Communication from Gayatri Vidya Parishad College of Engineering, Visakhapatnam, India, in 2014 and an Master's in Computer Engineering in 2019 from the University of North Texas, Denton, USA. He is currently a Ph.D. student in the research group at Smart Electronics Systems Laboratory (SESL) at Computer Science and Engineering at the University of North Texas, Denton, TX. His research interests include smart healthcare and Blockchain applications in Internet of Things (IoT). 
\end{IEEEbiography}

\vspace{-1.0cm}

\begin{IEEEbiography}
[{\includegraphics[height=1.3in,keepaspectratio]{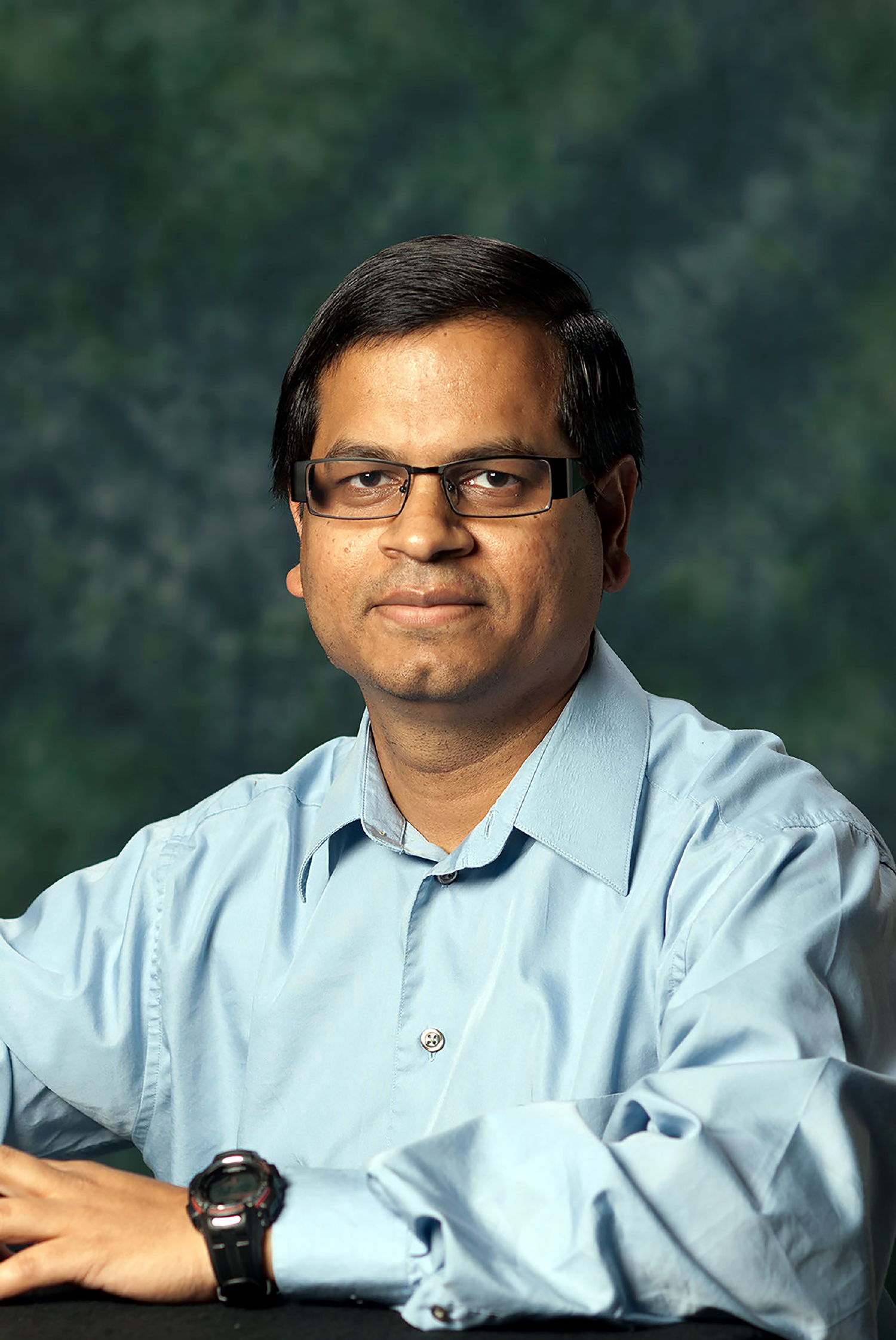}}] 
{Saraju P. Mohanty} received the bachelor’s degree (Honors) in electrical engineering from the Orissa University of Agriculture and Technology, Bhubaneswar, in 1995, the master’s degree in Systems Science and Automation from the Indian Institute of Science, Bengaluru, in 1999, and the Ph.D. degree in Computer Science and Engineering from the University of South Florida, Tampa, in 2003. He is a Professor with the University of North Texas. His research is in ``Smart Electronic Systems'' which has been funded by National Science Foundations (NSF), Semiconductor Research Corporation (SRC), U.S. Air Force, IUSSTF, and Mission Innovation. He has authored 400 research articles, 4 books, and 7 granted and pending patents. His Google Scholar h-index is 45 and i10-index is 180 with 8600 citations. He is regarded as a visionary researcher on Smart Cities technology in which his research deals with security and energy aware, and AI/ML-integrated smart components. He introduced the Secure Digital Camera (SDC) in 2004 with built-in security features designed using Hardware-Assisted Security (HAS) or Security by Design (SbD) principle. He is widely credited as the designer for the first digital watermarking chip in 2004 and first the low-power digital watermarking chip in 2006. He is a recipient of 14 best paper awards, Fulbright Specialist Award in 2020, IEEE Consumer Technology Society Outstanding Service Award in 2020, the IEEE-CS-TCVLSI Distinguished Leadership Award in 2018, and the PROSE Award for Best Textbook in Physical Sciences and Mathematics category in 2016. He has delivered 15 keynotes and served on 13 panels at various International Conferences. He has been serving on the editorial board of several peer-reviewed international transactions/journals, including IEEE Transactions on Big Data (TBD), IEEE Transactions on Computer-Aided Design of Integrated Circuits and Systems (TCAD), IEEE Transactions on Consumer Electronics (TCE), ACM Journal on Emerging Technologies in Computing Systems (JETC), and Springer Nature Cpmputer Science (SN-CS). He was the Editor-in-Chief of the IEEE Consumer Electronics Magazine (MCE) during 2016-2021. He served as the Chair of Technical Committee on Very Large Scale Integration (TCVLSI), IEEE Computer Society (IEEE-CS) during 2014-2018 and on the Board of Governors of the IEEE Consumer Electronics Society during 2019-2021. He serves on the steering, organizing, and program committees of several international conferences. He is the steering committee chair/vice-chair for the IEEE International Symposium on Smart Electronic Systems (IEEE-iSES), the IEEE-CS Symposium on VLSI (ISVLSI), and the OITS International Conference on Information Technology (OCIT). He has mentored 2 post-doctoral researchers, and supervised 13 Ph.D. dissertations, 26 M.S. theses, and 12 undergraduate projects.
\end{IEEEbiography}

\vspace{-1.0cm}

\begin{IEEEbiography}
[{\includegraphics[height=1.3in,keepaspectratio]{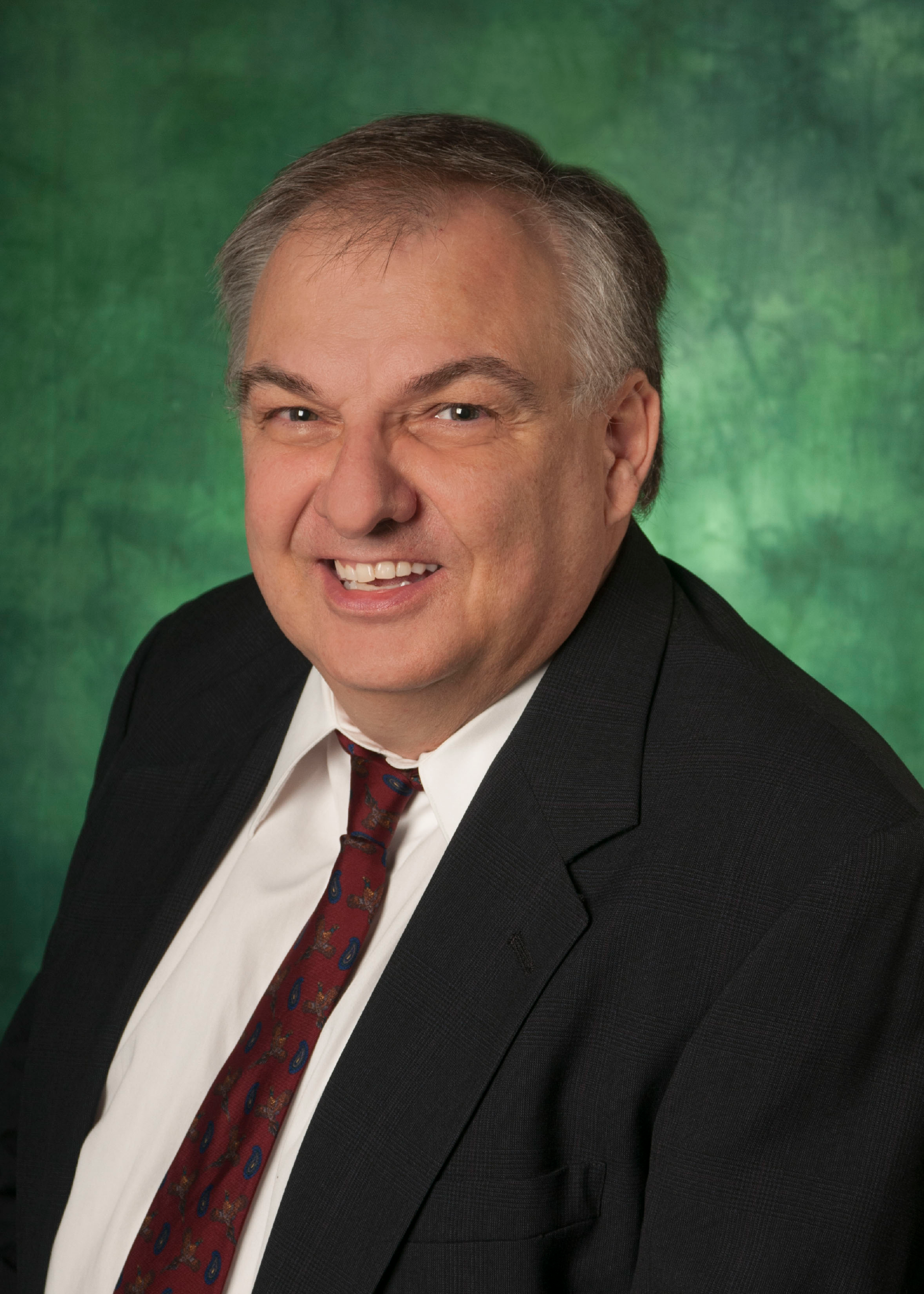}}]
{Elias Kougianos} received a BSEE from the University of Patras, Greece in 1985 and an MSEE in 1987, an MS in Physics in 1988 and a Ph.D. in EE in 1997, all from Louisiana State University. 
From 1988 through 1998 he was with Texas Instruments, Inc., in Houston and Dallas, TX. 
In 1998 he joined Avant! Corp. (now Synopsys) in Phoenix, AZ as a Senior Applications engineer and in 2000 he joined Cadence Design Systems, Inc., in Dallas, TX as a Senior Architect in Analog/Mixed-Signal Custom IC design. He has been at UNT since 2004. He is a Professor in the Department of Electrical Engineering, at the University of North Texas (UNT), Denton, TX. His research interests are in the area of Analog/Mixed-Signal/RF IC design and simulation and in the development of VLSI architectures for multimedia applications. 
He is an author of over 140 peer-reviewed journal and conference publications.  
\end{IEEEbiography}

\end{document}